\documentclass[fleqn,usenatbib]{mnras}

\usepackage{newtxtext,newtxmath}

\usepackage[T1]{fontenc}
\usepackage{ae,aecompl}

\usepackage{graphicx}	
\usepackage{amsmath}	
\usepackage{amssymb}	
\usepackage{multirow}
\usepackage{txfonts}
\usepackage{amsmath}
\usepackage{float}

\newcommand{\teff}{$T_\mathrm{eff}$}
\newcommand{\logg}{$\log g$}
\newcommand{\feh}{[Fe/H]}
\newcommand{\sife}{[Si/Fe]}
\newcommand{\mgfe}{[Mg/Fe]}

\newcommand{\mic}{$\mu \mathrm m$}

\title[Inner Bulge]{Chemical Characterization of the Inner Galactic bulge:\\ North-South  Symmetry\thanks{Based on observations collected at the European Southern Observatory, Chile, program numbers 089.B-0312(A)/VM/CRIRES, 089.B-0312(B)/VM/ISAAC, 091.B-0369(A)/VM/SOFI, and 091.B-0369(B)/SM/CRIRES.}}

\author[G. Nandakumar et al.]{%
G. Nandakumar$^{1}$\thanks{E-mail: Govind.Nandakumar@oca.eu},
N. Ryde$^{2}$,
M. Schultheis$^{1}$,
B. Thorsbro$^{2}$,
H. J\"onsson$^{2}$,
\newauthor%
P. S. Barklem$^{3}$,
R. M. Rich$^{4}$,
and F. Fragkoudi$^{5,6}$
\\
$^{1}$Laboratoire Lagrange, Universit\'e C\^ote d'Azur, Observatoire de la C\^ote d'Azur, CNRS, Blvd de l'Observatoire, F-06304 Nice, France\\
$^{2}$Lund Observatory, Department of Astronomy and Theoretical Physics, Lund University, Box 43, SE-221 00 Lund, Sweden\\
$^{3}$ Theoretical Astrophysics, Department of Physics and Astronomy, Uppsala University, Box 516, 751 20 Uppsala, Sweden\\
$^{4}$ Department of Physics and Astronomy, UCLA, 430 Portola Plaza, Box 951547, Los Angeles, CA 90095-1547, USA\\
$^{5}$ Max-Planck-Institut f\"ur Astrophysik, Karl-Schwarzschild-Str. 1, 85741 Garching, Germany\\
$^{6}$ GEPI, Observatoire de Paris, PSL Research University, CNRS, Sorbonne Paris Cité, 5 place Jules Janssen, 92190 Meudon, France
}

\date{Accepted XXX. Received YYY; in original form ZZZ}

\pubyear{2015}

\begin{document}
\label{firstpage}
\pagerange{\pageref{firstpage}--\pageref{lastpage}}
\maketitle

\begin{abstract}

While the number of stars in the Galactic bulge with detailed chemical abundance measurements is increasing rapidly, the inner Galactic bulge ($\rm |b| < 2\degr$) remains poorly studied, due to heavy interstellar absorption and photometric crowding. We have carried out a high-resolution IR spectroscopic study of 72 M giants in the inner bulge using the CRIRES (ESO/VLT) facility. Our spectra cover the wavelength range of 2.0818 -- 2.1444 $\mu$m with the resolution of R$\sim$50,000 and have signal-to-noise ratio of 50--100. Our stars are located along the bulge minor axis at l = 0$\degr$, b = $\pm$0\degr, $\pm$1\degr, $\pm$2\degr and +3\degr. Our sample was analysed in a homogeneous way using  the most current K-band line list. We clearly detect a bimodal MDF with a metal-rich peak at $\rm \sim +0.3\,dex$ and a metal-poor peak at $\rm \sim -0.5\,dex$, and no stars with \feh\ $>$ +0.6\,dex. The Galactic Center field reveals  in contrast a mainly metal-rich population with a mean metallicity of $\rm +0.3\,dex$. We derived \mgfe\ and \sife\ abundances which are consistent with trends from the outer bulge. We confirm for the supersolar metallicity stars the decreasing trend in \mgfe\ and \sife\ as expected from chemical evolution models. With the caveat of a relatively small sample, we do not find significant differences in the chemical abundances between the Northern and the Southern fields, hence the evidence is consistent with symmetry in chemistry between North and South.   

\end{abstract}




\section{Introduction}

The Milky Way bulge is on average one of the older components of the Milky Way, located within $\sim 3-4$\,kpc from the Galactic center and harbors most likely multiple stellar populations. Its structure, formation, and evolution scenarios have been under debate in recent years owing to large increase of observational data allowed by the advancements in telescope sizes, instrumentation as well as observing techniques. 

The Galactic bulge formation from the dissipative collapse of a primordial gas cloud \citep{eggen:62} or through hierarchical mergers in lambda cold CDM (\citealt{scannapieco:03,abadi:03,immeli:04}) leads to a centrally concentrated spheroidal structure or classical bulge as seen in simulations of Galaxy formation \citep{abadi:03}. But the boxy/peanut shaped (B/P) or X-shaped morphology of the bulge stellar populations (red clump stars in particular) revealed from photometric surveys (COBE/DIRBE: \citealt{weiland:94}, 2MASS: \citealt{mcwilliam:94}, OGLE-III: \citealt{nataf:10}, VVV: \citealt{wegg:13,valenti:16}) suggest another scenario wherein the bulge forms from the secular internal evolution of the early disc. This leads to the bar formation, which subsequently buckles and redistributes the disc angular momentum in the disc. The cylindrical rotation exhibited by bulge stars in the outer bulge fields (BRAVA: \citealt{kunder:12}, ARGOS: \citealt{ness:13a}, GIBS: \citealt{zoccali:14}) have been successfully reproduced using N-body simulations of pure thin-disc models \citep{shen:10,valpuesta:11,gardner:14,dimatteo:15} as the result of secular evolution of disc. Thus, this scenario is able to explain not only the morphology but also the chemical and kinematic properties of bulge stars with \feh\,$>$-0.5\,dex to an extent, whereas it is unable to account for the metal-poor bulge (\feh\,$<$-0.5\,dex) stars that shows an extended, centrally concentrated spheroidal distribution \citep{GIBSIII}. However,  \citet{dimatteo:14,dimatteo:15,dimatteo:16,fragkoudi:17} have suggested using N-body simulations that the Milky Way bulge is the result of the mapping of the "thin + thick" disc of the Galaxy into the boxy/peanut-shaped structure and thus also explains the origin of metal-poor stars to arise from the thick disc metal-poor population (and see also \citet{athanassoula_merge:16,debattista:17} for similar findings using hydrodynamic simulations). For a detailed review of the Bulge formation scenarios in the context of chemical, dynamical, chemodynamical and cosmological models, see \citet{Barbuy:2018}.


These general advances derive from observations of bulge stars in regions of relatively low optical extinction. However, much of the {\it mass} of the bulge, that lies within $\sim$400-500 pc around the Galactic center or within $|b|\sim 3^\circ$, has escaped investigation due to the obscuration in the line-of-sight toward the inner bulge. Observations at infrared wavelengths mitigates this problem, enabling us to peer further and deeper through the highly extincted inner regions of the Milky Way. 

One of the first observations and metallicity determination for stars in the inner bulge region along the bulge minor axis ($0\fdg1 <\rm b< -2\fdg8$) was carried out by \cite{frogel:99} using IRCAM \citep{persson:92}. These authors derived mean metallicities from the slope of giant branches in CMDs and estimated a slope of -0.085$\pm$0.033\,dex\,deg$^{-1}$ for -0\fdg8$\leq$b$\leq$-2\fdg8. Follow up low resolution (R$\sim$1300-4800) spectroscopic observations of around 100 M giants in the same fields were carried out by \cite{ramirez:00}. These authors determined metallicities for their sample from the equivalent widths of three features in these spectra, EW(Na), EW(Ca), and EW(CO), and found no evidence for a metallicity gradient along the minor or major axes of the inner bulge ($R_{G}$ $\sim$560\,pc). \cite{carr2000} and \mbox{\cite{ramirez2000}}	 have carried out a detailed abundance analysis using high resolution spectra (R$\sim$40,000) to estimate a mean metallicity of $+$0.12$\pm$0.22\,dex for 10 cool luminous super giants stars in the Galactic center (GC). The same stars were re-analysed by \cite{cunha2007} with a slightly higher resolution (R$\sim$50,000) and they estimated a similar mean metallicity. 

\citet{rich:2007} did one of the first detailed abundance analysis of 17 M giants located at ($l$,$b$)=($0^\circ,-1^\circ$) from their high resolution (R$\sim$25,000) spectra and found a mean iron abundance of [Fe/H]=$-0.22$ with a $\rm 1\,\sigma$ dispersion of 0.14\,dex. \citet{rich:12} then proceeded to carry out a consistent analysis of 30 M giants at ($l$,$b$)=($0,-1\fdg75$) and ($1^\circ,-2\fdg75$) that were observed with the same instrument and estimated mean iron abundances of $-0.16\pm0.12$\,dex and $-0.21\pm0.08$\,dex respectively. These authors also combined their analysis of 14 M giants in Baade's Window (l=$1\fdg02$, $b$=$-3\fdg93$) using the same instrument and found no major vertical abundance or abundance gradient in the inner most 150\,pc to 600\,pc region. \cite{babusiaux:14} determined metallicities for $\sim$100 red clump (RC) stars at ($l$,$b$) = ($0^\circ$,+$1^\circ$) using low resolution spectra (R$\sim$6,500) and found a mean metallicity very similar to that of \cite{ramirez:00} and \cite{rich:07} at ($l$,$b$) = ($0^\circ, -1^\circ$) suggesting symmetry between Northern and Southern inner bulge fields. APOGEE data \citep{majewski:17} presented evidence of a significant metal-poor component in the inner one degree \citep{schultheis:15}. \citet{ryde_schultheis:15} analyzed 9 field giants in the vicinity of the Nuclear Star Cluster (NSC), finding a broad, metal-rich component at [Fe/H]$\sim$+0.1 and a lack of metal-poor stars. Their $\alpha$ abundances are found to be low following the trends from studies in the outer bulge, resembling a bar-like population. \citet{grieco:15} made a detailed chemical evolution model of the center region to compare with these data. They conclude that this region experienced a main, early, strong burst of star-formation, with a high star-formation efficiency, no late in-fall of gas, and some indications of a top-heavy Initial-Mass Function (IMF). \citet{ryde:16} presented an abundance study of 28 M giants in fields located within a few degrees south of the Galactic Center using high resolution ($R\sim$50,000) spectra and determine the metallicity distribution functions and $\alpha$ element trends. They find a large range of metallicities, with a narrower range towards the center, and a large similarity of the $\alpha$ element trends among the different fields and that of the outer bulge, suggesting a homogeneous bulge regarding the enrichment processes and star-formation history. Recently, \cite{GIBSIII}  derived metallicities of stars from the GIRAFFE Inner Bulge Survey (GIBS) in fields close to the Galactic plane, at $b$ = -2\degr and $b$ = -1\degr. They find a clear bimodal Metallicity Distribution Function (MDF). Meanwhile, \cite{garciaperez:18} identify more than two components by arranging the red giant stars in the APOGEE DR12 bulge sample by projected Galactocentric distance and distance from the Galactic mid-plane. \cite{fragkoudi:18}, using derived metallicities from the more recent data release of APOGEE (DR13), compare the observed MDF, including that in the inner bulge with that obtained from N-body simulations of a composite (thin+thick) stellar disc. These authors find the MDF trends to be reproduced by the models, and argue their consistency with  bulge formation from the secular evolution of disc.  	 					


  It is clear from the overview of the literature that almost all focus has been on the less extincted Southern fields of the inner bulge. Here, we present the first comprehensive data set of high resolution spectroscopic data of 72 stars in fields sampling the full inner bulge within $|b|\sim 3^\circ$, thus both at Southern and Northern latitudes, all analyzed in a homogeneous way. Our study is the first comparative study of stars in inner fields of the bulge of both sides of the Galactic Plane. With this study, we aim to estimate the nature of the MDF in the inner bulge region and do a comparison to its nature in the outer bulge. We also intend for the first time to investigate the symmetry in MDF between Northern and Southern latitudes, which has not been studied before. With our limited sample, we also try to understand how the mean metallicity changes with different latitudes as we move away from the Galactic mid-plane. In addition, we also estimate $\alpha$-element abundances for our stars, and show the [$\alpha$/Fe] vs \feh\, trend in the inner bulge region. Finally, we have measured radial velocities of our stars with the aim of exploring correlations between kinematics and metallicity. All these studies are needed to constrain the formation of the Milky Way bulge.   


The paper is organized as follows. In Section 2 we describe the high and low resolution spectroscopic observations, our target selection and the instruments used. The detailed spectroscopic data analysis and stellar parameter determination are given in Section 3. We show our results in Section 4, with a discussion in Section 5 and conclusion in Section 7.  



\section{Observations}

\begin{figure}
    \includegraphics[width=\columnwidth]{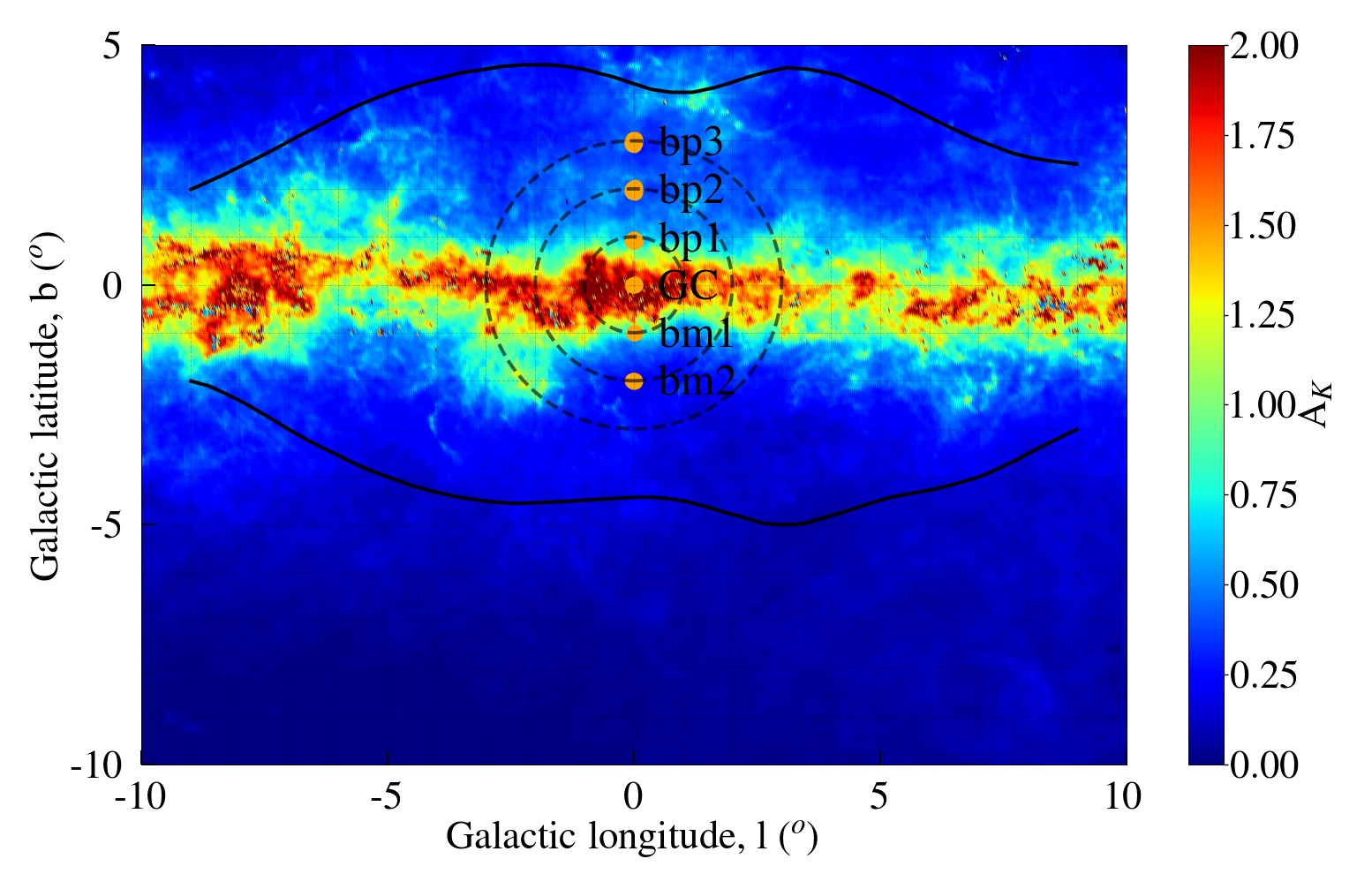}
    \caption{Our inner bulge fields in galactic coordinates. The individual stellar coordinates (orange) are plotted, which shows the concentration within each field. The mean extinction map, calculated from \protect\cite{gonzalez2012} using extinction coefficients from \protect\cite{nishiyama2009}, is overlaid on the figure to show the level of obscuration in our observed fields. The bulge outline is the COBE/DIRBE bulge envelope \citep{weiland:94}.}
    \label{fig:sky}
\end{figure}

\begin{figure}
	\includegraphics[width=\columnwidth]{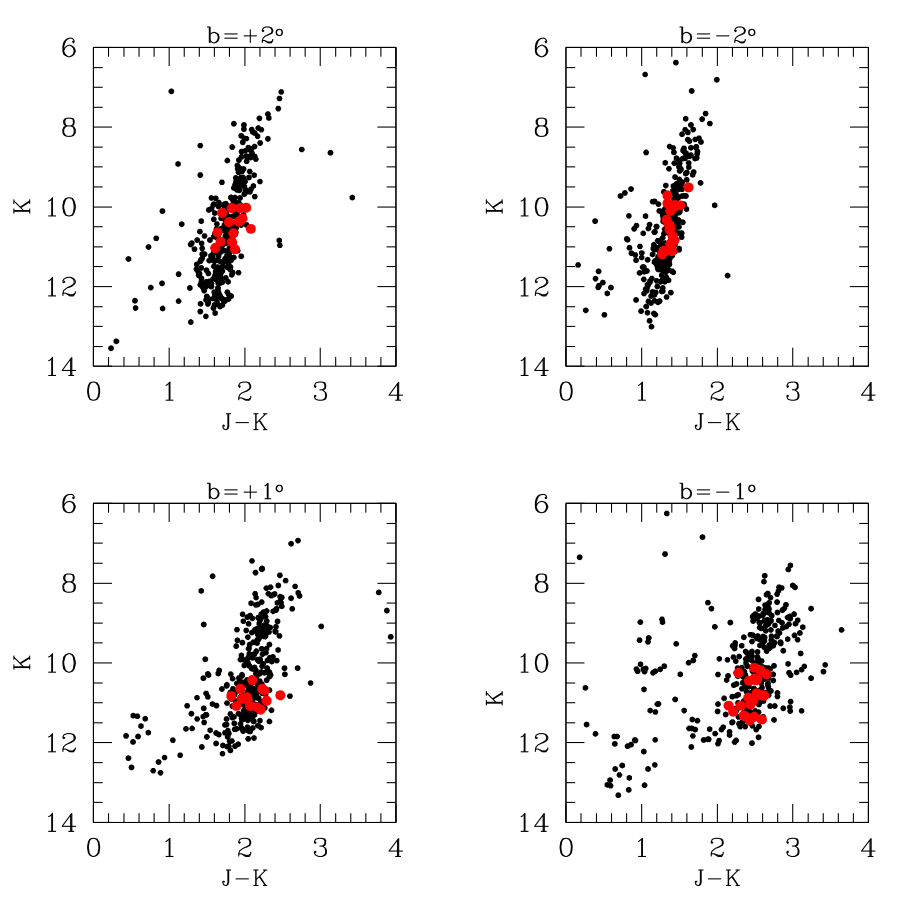}
    \caption{2MASS K vs . J--K diagram for the Northern and Southern fields. The filled red points show our selected targets.}
    \label{fig:cmd}
\end{figure}

In order to determine the MDF and the $\alpha$-element trends for stellar populations in the optically obscured inner bulge, we have observed M giants in six different fields at high spectral resolution in the K band. We succeeded in securing useful spectra of 72 bulge giants along the Northern and Southern minor axes within 3 degrees\footnote{This corresponds to approximately 450 pc in projected distance} of the Galactic Center. We have included a Galactic Center field at $2.5-5.5\arcmin$ North of the Galactic Center, thus avoiding the Nuclear Star Cluster\footnote{An abundance analysis based on high-resolution spectra of stars in the Nuclear Star Cluster is performed in \citet{ryde:16:nsc,rich:17}}. In Figure \ref{fig:sky} the stars are plotted on a sky projection showing how they sample the inner minor axis both to the North and to the South of the Galactic Plane. We observed the stars in the infrared K-band at $2.1$\,\mic, in order to overcome the extreme optical extinction towards the inner regions. 

All stars were observed at both high- and low spectral resolution. The high resolution spectra, recorded with the CRIRES spectrograph \citep[yielding  a resolving power of $R\sim 50,000$;][]{crires,crires2,crires1} mounted on UT1 of the {\it Very Large Telescope}, VLT, were used to determine the metallicity and abundances of the $\alpha$ elements, while the low resolution spectra, recorded with the ISAAC spectrograph \citep[$R\sim 1000$; ][]{isaac} on UT3 of the VLT and the SOFI instrument \citep[$R\sim 1000$][]{sofi} on the NTT telescope at La Silla, were used to determine the effective temperatures of the stars via low resolution spectroscopy of the 2.2 $\mu$m CO bands, following the method described in \citet{schultheis:16}.

44 of the observed giants that are located in the Northern fields were observed with CRIRES during the period 20 April - 16 September 2013 in Service Mode and with SOFI during 13-17 July 2013 in Visitor Mode (program ID 091.B-0369). 28 of the remaining giants are located in the Galactic Center field and the Southern fields and were observed with both the CRIRES and ISAAC spectrographs in Visitor Mode during 27-29 June 2012 (program ID 089.B-0312).  The 44 bulge giants in the Northern fields are presented for the first time here, whereas the 28 giants in Southern and Center fields are reanalysed here for consistency, but were first published in \citet{ryde_schultheis:15} and \citet{ryde:16}.  In Tables \ref{starsall_N_tab} and \ref{starsall_S_tab} the coordinates and total exposure times for all the stars are provided. 

\subsection{Target Selection}

\begin{figure}
	\includegraphics[scale=0.5,width=\columnwidth]{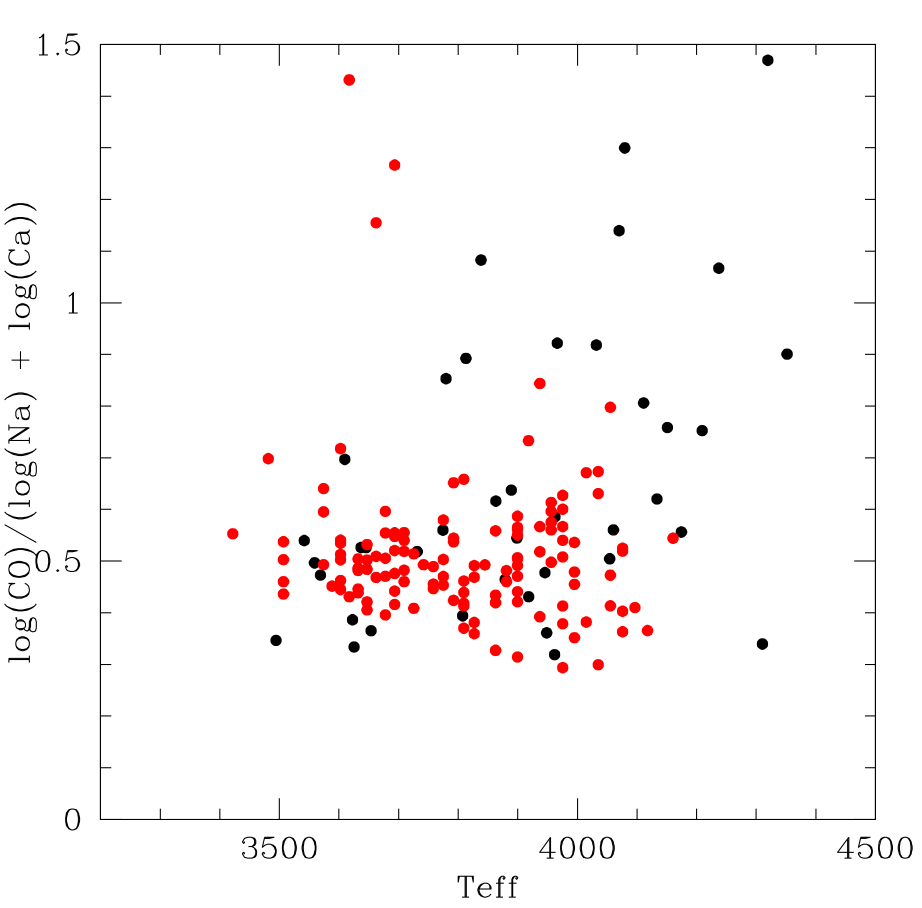}
    \caption{Effective temperatures based on the $\rm 12^{CO}$ first overtone bandhead vs. log\,g sensitive spectral index log(CO)/(log(Na)+log(Ca)). Black filled circles show our sample while the red filled circles are those of \citet{ramirez2000}.}
    \label{fig:example_figure}
\end{figure}

We selected  our M giant sample in the Northern fields in the same way as done for the Southern fields (\citealt{ryde:16}) to ensure consistency of the North and South samples.
We used the dereddened color-magnitude diagram, which covers the full
($J-K$)$_0$-colour range. Figure \ref{fig:cmd} shows the corresponding colour-magnitude diagrams for the northern and southern fields together with our selected targets.
Interstellar extinction is less severe than in the Southern fields, especially the field located at $b$=+$1^{\circ}$ is well known for its low and homogeneous interstellar extinction, which has been the subject of many  studies (see e.g. \citealt{omont2003}, \citealt{vanloon2003}, \citealt{babusiaux:14}). Our input catalog for the Northern field is the 2MASS catalog,  as the  \citet{nishiyama2009} catalog is not available for the Northern fields.
We again checked our extinction values with the 3D high-resolution interstellar extinction map from \citet{schultheis:14}. 

In addition, we used the surface-gravity index, as outlined in \citet{ramirez:00}, based on measured equivalent widths ($W$) of the Na\,{\sc i}- and  Ca\,{\sc i}-features, as well as the first overtone, CO bandhead in our low-resolution spectra. This is done to ensure that none of our stars actually is a foreground dwarf star. \citet{ramirez:00} demonstrated nicely that
this index is a very good discriminator to distinguish dwarf stars from giant  stars. Figure \ref{fig:example_figure} shows our sample of M giant stars in black and the comparison sample of \citet{ramirez1997} in red. Dwarf stars are supposed to lie at about $\log$ [((CO)/((Ca) + (Na))]$ \sim 0$. We see clearly that our stars are indeed M giants. 

\begin{figure}
    \includegraphics[width=\columnwidth]{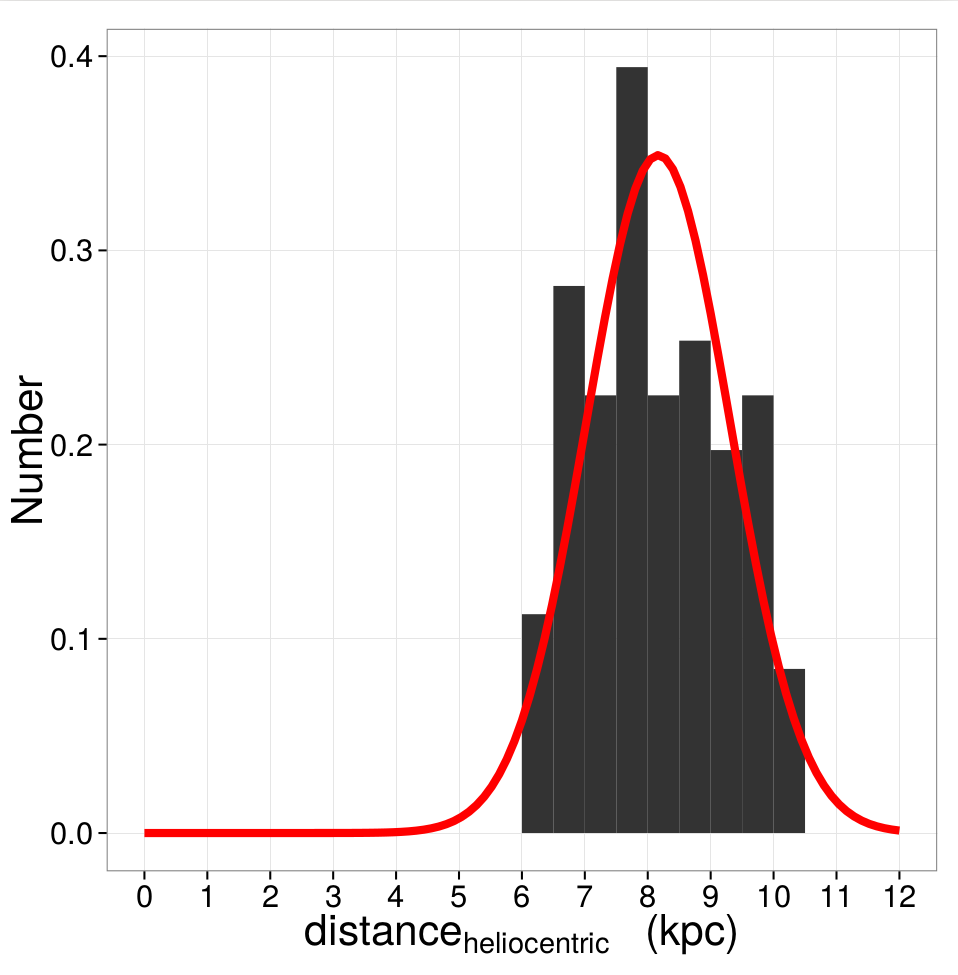}
    \caption{Histogram of the heliocentric distances of our stars using the isochrone method from \citet{rojas:15} and  \citet{schultheis:17}}
    \label{fig:dist}
\end{figure}

We have also calculated the heliocentric distances for our sample by
using the stellar parameters \teff, \logg\ and \feh\  (determined in Section \ref{params}) and taking the closest point
in the PARSEC isochrones \citep{bressan:12}, by assuming  a typical age between 5 and 10\,Gyr. A more detailed explanation of the procedure can be found in \citet{rojas:15} and \citet{schultheis:17}. Figure~\ref{fig:dist}
shows the histogram of our derived distances. Our distance distribution  shows clearly that our stars
belong to the bulge  and that we can exclude any foreground contamination.

All our targeted stars are M giants (3300\,K$<$ \teff $<$4300\,K and  0.5 $< \log g <$ 2.0), for which we will determine the metallicities, [Fe/H], and the abundances of the $\alpha$-elements, Mg and Si. The apparent magnitudes of our stars lie in the range of $K$=$ 9.5-12.0$. The $H$ and $K$-band magnitudes, as well as the dereddened $H_0$- and $K_0$-magnitudes are presented in Table \ref{starsall_N_tab} and \ref{starsall_S_tab}.

Following, we give some details about the specific observations of our targets.

\begin{table*}
\caption{Basic data for the observed stars in the Northern fields and their stellar parameters.}\label{starsall_N_tab}
\begin{tabular}{l c c c c c c c c c c c c c}
\hline
\hline
\multicolumn{1}{l}{Star} & RA (J2000) & Dec (J2000) & $H$ & $K$ & $H_0$ & $K_0$  &\teff & \logg & \feh &  $\xi_\mathrm{macro}$ & \sife & \mgfe   & exp. time \\
         &    (h:m:s)    & (d:am:as)   &   &  &  &  &  [K] & (cgs) &   & [km\,s$^{-1}$] &  & [s] \\         
\hline
\multicolumn{4}{l}{Northern field at $(l,b) = (0^{\circ},+3^{\circ})$}\\
\hline
bp3-01 & 17:34:05.20 & -27:19:58.20 & 10.04 & 9.55 & 9.31 & 9.10 & 3780 & 0.79 & $-$0.60 & 6.7 & $+$0.20 & $+$0.20 & 1500\\
bp3-02 & 17:34:19.24 & -27:20:42.70 & 11.48 & 10.99 & 10.69 & 10.50 & 4111 & 1.90 & $+$0.20 & 5.6 & $-$0.12 & $-$0.10 & 1200\\
bp3-04 & 17:34:09.23 & -27:22:38.00 & 10.97 & 10.39 & 10.22 & 9.23 & 3623 & 0.94 & $+$0.10 & 7.3 & $+$0.05 & 0.00 & 1200\\
bp3-05 & 17:34:00.71 & -27:18:59.60 & 10.47 & 9.96 & 9.78 & 9.53 & 3879 & 1.55 & $+$0.30 & 5.4 & $-$0.20 & $-$0.10 & 1200\\
bp3-06 & 17:34:07.13 & -27:22:28.90 & 11.71 & 11.21 & 10.97 & 10.75 & 3755 & 0.68 & $-$0.70 & 5.8 & $+$0.20 & $+$0.30 & 1200\\
bp3-07 & 17:34:24.09 & -27:23:16.70 & 11.52 & 10.97 & 10.77 & 10.51 & 3962 & 1.70 & $+$0.30 & 4.9 & $-$0.10 & 0.00 & 1200\\
bp3-08 & 17:34:16.44 & -27:24:34.30 & 10.43 & 9.82 & 9.69 & 9.36 & 3637 & 1.12 & $+$0.30 & 4.9 & $-$0.20 & 0.00 & 1200\\ 
bp3-10 & 17:34:13.70 & -27:24:31.00 & 11.50 & 11.03 & 10.75 & 10.57 & 4052 & 1.70 & $+$0.10 & 5.3 & $-$0.10 & $+$0.10 & 1200\\
bp3-11 & 17:33:58.67 & -27:20:30.20 & 10.10 & 9.59 & 9.44 & 9.18 & 3542 & 0.66 & $+$-0.10 & 6.8 & $-$0.10 & 0.00 & 720\\ 
bp3-12 & 17:34:06.61 & -27:19:29.30 & 10.13 & 9.61 & 9.40 & 9.16 & 3966 & 1.45 & $-$0.05 & 7.2 & 0.00 & $+$0.10 & 720\\
bp3-13 & 17:34:08.21 & -27:22:58.70 & 10.25 & 9.66 & 9.50 & 9.20 & 3625 & 0.74 & $-$0.20 & 5.2 & $-$0.05 & 0.00 & 720\\
bp3-14 & 17:33:57.31 & -27:21:07.60 & 10.27 & 9.67 & 9.61 & 9.26 & 3569 & 0.92 & $+$0.20 & 9.2 & $-$0.20 & $-$0.10 & 1200\\
bp3-15 & 17:34:06.52 & -27:22:30.20 & 10.30 & 9.77 & 9.56 & 9.31 & 3610 & 0.56 & $-$0.50 & 6.3 & $+$0.20 & $+$0.10 & 1200\\
bp3-16 & 17:34:15.46 & -27:22:30.90 & 10.39 & 9.78 & 9.63 & 9.31 & 3654 & 0.85 & $-$0.10 & 8.4 & $+$0.10 & $+$0.20 & 1200\\
bp3-17 & 17:34:03.57 & -27:18:49.60 & 10.31 & 9.82 & 9.60 & 9.38 & 3946 & 1.52 & $+$0.10 & 7.5 & $-$0.10 & $-$0.10 & 1200\\
\hline
\multicolumn{4}{l}{Northern field at $(l,b) = (0^{\circ},+2^{\circ})$}\\
\hline
bp2-01 & 17:37:49.64 & -27:51:05.50 & 11.21 & 10.66 & 10.42 & 10.17 & 4054 & 1.95 & $+$0.40 & 6.1 & $-$0.10 & $-$0.05 & 1800\\
bp2-02 & 17:38:02.25 & -27:52:31.40 & 11.11 & 10.64 & 10.33 & 10.16 & 3838 & 1.60 & $+$0.50 & 5.9 & $-$0.20 & $-$0.10 & 1800\\
bp2-03 & 17:37:58.65 & -27:51:40.90 & 10.93 & 10.37 & 10.15 & 9.89 & 3889 & 1.70 & $+$0.60 & 7.4 & $-$0.10 & $-$0.25 & 1200\\
bp2-04 & 17:38:02.12 & -27:52:49.80 & 10.86 & 10.38 & 10.08 & 9.90 & 4134 & 1.74 & $-$0.10 & 6.5 &  0.00 & $+$0.05 & 1800\\
bp2-05 & 17:37:53.31 & -27:52:01.10 & 11.39 & 10.88 & 10.60 & 10.39 & 4079 & 1.40 & $-$0.50 & 4.9 & $+$0.20 & $+$0.40 & 1800\\
bp2-06 & 17:37:51.19 & -27:54:01.70 & 11.48 & 11.03 & 10.72 & 10.56 & 3962 & 1.17 & $-$1.80 & 7.8 & $+$0.20 & $+$0.40 & 1200\\
bp2-07 & 17:37:58.13 & -27:53:42.60 & 11.30 & 10.87 & 10.55 & 10.40 & 4320 & 1.84 & $-$0.50 & 5.8 & $+$0.10 & $+$0.10 & 1800\\
bp2-08 & 17:37:56.88 & -27:51:52.80 & 10.87 & 10.29 & 10.08 & 9.80 & 3808 & 1.38 & $+$0.25 & 4.0 &  $-$0.15 & $+$0.20 & 1200\\
bp2-09 & 17:37:50.52 & -27:54:24.60 & 10.89 & 10.36 & 10.14 & 9.89 & 3813 & 0.70 & $-$0.80 & 6.2 & $+$0.15 & $+$0.20 & 1200\\
bp2-10 & 17:38:01.25 & -27:55:38.40 & 10.60 & 10.03 & 9.86 & 9.57 & 3774 & 1.28 & $+$0.20 & 4.8 & 0.00 & 0.00 & 1800\\
bp2-11 & 17:38:00.03 & -27:55:45.00 & 10.62 & 10.02 & 9.88 & 9.56 & 3645 & 1.26 & $+$0.50 & 5.3 & $-$0.30 & $+$0.10 & 1200\\
bp2-12 & 17:38:02.01 & -27:53:30.70 & 10.55 & 10.03 & 9.78 & 9.56 & 4032 & 1.54 & $-$0.10 & 8.4 & 0.00 & 0.00 & 2400\\
bp2-13 & 17:38:02.54 & -27:55:42.50 & 11.52 & 11.06 & 10.78 & 10.60 & 3899 & 1.12 & $-$0.40 & 5.5 & $+$0.15 & $+$0.30 & 1200\\
bp2-14 & 17:38:01.74 & -27:55:56.10 & 11.16 & 10.55 & 10.42 & 10.09 & 3559 & 1.00 & $+$0.30 & 5.4 & $-$0.10 & 0.00 & 1200\\
bp2-15 & 17:38:13.55 & -27:53:41.40 & 10.67 & 10.15 & 9.93 & 9.69 & 4237 & 1.69 & $-$0.50 & 6.7 & $+$0.13 & $+$0.20 & 600\\
\hline
\multicolumn{4}{l}{Northern field at $(l,b) = (0^{\circ},+1^{\circ})$}\\
\hline
bp1-01 & 17:41:57.26 & -28:28:46.50 & 11.91 & 11.17 & 10.87 & 10.53 & 3918 & 1.54 & $+$0.20 & 5.6 & 0.00 & 0.00 & 1200\\
bp1-02 & 17:41:57.62 & -28:28:51.60 & 11.46 & 10.86 & 10.41 & 10.21 & 4311 & 1.89 & $-$0.40 & 6.0 & $+$0.10 & $+$0.10 & 1200\\
bp1-03 & 17:41:58.61 & -28:28:46.80 & 11.38 & 10.65 & 10.31 & 9.99 & 3948 & 1.75 & $+$0.40 & 4.8 & $-$0.10 & $-$0.20 & 1800\\ 
bp1-04 & 17:41:55.92 & -28:27:03.80 & 11.56 & 10.93 & 10.60 & 10.34 & 4060 & 1.66 & $+$0.00 & 6.6 & $-$0.15 & 0.00 & 1200\\
bp1-05 & 17:41:59.76 & -28:27:36.20 & 11.70 & 10.95 & 10.68 & 10.32 & 3731 & 1.30 & $+$0.30 & 4.6 & $-$0.10 & $+$0.20 & 1200\\
bp1-06 & 17:41:57.66 & -28:27:07.20 & 11.79 & 11.11 & 10.82 & 10.51 & 3863 & 1.50 & $+$0.30 & 7.4 & $-$0.06 & $-$0.10 & 1200\\ 
bp1-07 & 17:42:03.59 & -28:27:18.20 & 11.68 & 11.08 & 10.63 & 10.43 & 4209 & 1.58 & $-$0.60 & 5.9 & $+$0.10 & $+$0.30 & 1200\\ 
bp1-08 & 17:41:58.51 & -28:26:18.70 & 11.61 & 10.81 & 10.53 & 10.15 & 4242 & 1.63 & $-$0.60 & 7.2 & $+$0.14 & $+$0.30 & 1800\\
bp1-09 & 17:42:04.43 & -28:26:55.60 & 11.61 & 10.81 & 10.58 & 10.17 & 3494 & 0.79 & $+$0.20 & 5.2 & $-$0.10 & $-$0.15 & 1800\\
bp1-10 & 17:42:06.49 & -28:27:09.10 & 11.43 & 10.69 & 10.36 & 10.03 & 3879 & 1.50 & $+$0.20 & 5.7 & $-$0.13 & $+$0.10 & 1800\\
bp1-11 & 17:41:59.06 & -28:26:03.90 & 11.64 & 11.08 & 10.70 & 10.50 & 4352 & 1.90 & $-$0.50 & 7.3 & $+$0.18 & $+$0.20 & 1200\\
bp1-12 & 17:41:58.72 & -28:25:58.80 & 11.50 & 10.90 & 10.56 & 10.32 & 4175 & 1.95 & $+$0.10 & 6.4 & 0.00 & $+$0.05 & 1800\\
bp1-13 & 17:42:05.96 & -28:26:59.10 & 11.03 & 10.43 & 9.98 & 9.78 & 4151 & 1.47 & $-$0.60 & 6.9 & $+$0.10 & $+$0.20 & 1200\\
bp1-14 & 17:42:10.05 & -28:25:36.90 & 11.47 & 10.83 & 10.48 & 10.22 & 4070 & 1.70 & $+$0.00 & 5.4 & $-$0.06 & $+$0.10 & 1800\\
\hline
\hline
\end{tabular}
\end{table*}

\begin{table*}
\caption{Basic data for the observed stars in the Southern fields and their stellar parameters.}\label{starsall_S_tab}
\begin{tabular}{l c c c c c c c c c c c c c}
\hline
\hline
\multicolumn{1}{l}{Star} & RA (J2000) & Dec (J2000) & $H$ & $K$ & $H_0$ & $K_0$  &\teff & \logg & \feh &  $\xi_\mathrm{macro}$   & \sife & \mgfe   & exp. time \\
         &    (h:m:s)    & (d:am:as)   &   &  &  &  &  [K] & (cgs) &   & [km\,s$^{-1}$] & & & [s] \\         
\hline
\multicolumn{4}{l}{Galactic Centre field at $(l,b) = (0^{\circ},0^{\circ})$}\\
\hline
GC1   & 17:45:35.43 &  -28:57:19.28 & 14.74 & 11.90  &  10.48 & 9.40  & 3668 & 1.25 & $+$0.40   & 5.0 & $-$0.17 & $+$0.10 & 3000  \\
GC20  & 17:45:34.95 &  -28:55:20.17 & 14.47 & 11.87  &  10.21 & 9.37  & 3683 & 1.33 & $+$0.50   & 7.3 & $-$0.17 & $-$0.10 & 3600   \\
GC22  & 17:45:42.41 &  -28:55:52.99 & 13.41 & 11.54  &  9.15 & 9.04   & 3618 & 1.00 & $+$0.20   & 5.5 & $-$0.10 & $+$0.10 & 3600   \\
GC25  & 17:45:36.34 &  -28:54:50.41 & 14.35 & 11.60  &  10.09 & 9.10  & 3340 & 0.54 & $+$0.20   & 7.6 & $-$0.20 & $-$0.20 & 2400    \\
GC27  & 17:45:36.72 &  -28:54:52.37 & 14.31 & 11.64  &  10.05 & 9.14  & 3404 & 0.64 & $+$0.20   & 7.2 & $-$0.05 & $-$0.05 & 3600      \\
GC28  & 17:45:38.13 &  -28:54:58.32 & 14.08 & 11.67  &  9.82 & 9.17   & 3773 & 1.20 & $+$0.10   & 5.9 & $-$0.10 & $+$0.20 & 3000       \\
GC29  & 17:45:43.12 &  -28:55:37.10 & 14.39 & 11.59  &  10.13 & 9.09  & 3420 & 0.65 & $+$0.18   & 8.1 & $-$0.25 & $-$0.10 & 3600       \\
GC37  & 17:45:35.94 &  -28:58:01.43 & 13.77 & 11.50  &  9.51 & 9.00   & 3754 & 1.33 & $+$0.30   & 7.2 & $-$0.08 & $-$0.05 & 3600     \\
GC44  & 17:45:35.95 &  -28:57:41.52 & 13.78 & 11.74  &  9.52 & 9.24   & 3465 & 0.89 & $+$0.40   & 7.5 & $-$0.20 & $-$0.10 & 3600     \\
\hline
\multicolumn{4}{l}{Southern field at $(l,b) = (0^{\circ},-1^{\circ})$}\\
\hline
bm1-06 & 17:49:33.42 & -29:27:28.75 & 11.84 & 11.04 & 10.58 & 10.26 &  3814 & 	1.56 & 	 $+$0.50  & 4.6	 & $-$0.30 & $-$0.20 & 1800 \\
bm1-07 & 17:49:34.58 & -29:27:14.82 & 12.22 & 11.44 & 10.95 & 10.66 &  3873 & 	1.38 & 	 $+$0.08  & 5.7	 & $-$0.06 & $+$0.10 & 2400 \\
bm1-08 & 17:49:34.34 & -29:26:57.98 & 11.22 & 10.40 &  9.95 &  9.62 &  3650 & 	1.10 & 	 $+$0.24  & 4.2 & $-$0.05 & $-$0.05 & 1200 \\
bm1-10 & 17:49:34.45 & -29:26:48.68 & 10.92 & 10.12 &  9.65 &  9.34 &	3787 & 	1.09 & 	$-$0.10  & 7.3 & $+$0.05 & $+$0.20 & 1920 \\
bm1-11 & 17:49:32.57 & -29:26:30.75 & 11.26 & 10.41 & 10.00 &  9.63 &  3812 & 	1.35 & 	 $+$0.20  & 4.1	 & $-$0.10 & 0.00 & 1200 \\
bm1-13 & 17:49:37.12 & -29:26:40.24 & 10.91 & 10.25 &  9.65 &  9.47 &	3721 & 	0.46 & 	$-$0.91  & 6.1 & $+$0.20 & $+$0.50 & 1080 \\
bm1-17 & 17:49:37.08 & -29:26:21.67 & 11.70 & 10.94 & 10.43 & 10.16 &	3775 & 	0.65 & 	$-$0.79  & 7.3 & $+$0.25 & $+$0.40 & 1800 \\
bm1-18 & 17:49:37.83 & -29:26:19.19 & 12.23 & 11.42 & 10.96 & 10.64 &	3780 & 	1.28 & 	 $+$0.18 & 3.3  & $-$0.10 & 0.00 & 1800 \\	
bm1-19 & 17:49:36.93 & -29:26:10.51 & 12.11 & 11.34 & 10.84 & 10.56 &	3958 & 	1.77 & 	 $+$0.40  & 5.0 & $-$0.20 & $-$0.10 & 3600  \\
\hline
\multicolumn{4}{l}{Southern field at $(l,b) = (0^{\circ},-2^{\circ})$}\\
\hline
bm2-01 & 17:53:29.06 & -29:57:46.22 & 11.44 & 11.11 & 11.08 & 10.89  & 	3946 & 	1.33 & 	 $-$0.19 & 4.3 & $+$0.10 & $+$0.30 & 2400 \\
bm2-02 & 17:53:24.59 & -29:59:09.48 & 10.78 & 10.44 & 10.41 & 10.21  & 	4013 & 	1.51 & 	$-$0.10	 & 6.8 & $-$0.10 & $+$0.15 & 1920 \\
bm2-03 & 17:53:27.61 & -29:58:36.39 & 11.30 & 10.93 & 10.94 & 10.70  & 	3668 & 	1.25 & 	 $+$0.40 & 4.9 & 0.00 & --- & 2400 \\
bm2-05 & 17:53:33.20 & -29:57:25.88 & 10.07 &  9.51 &  9.70 &  9.28  &	 3450 & 0.64 & 	 $+$0.10 & 6.3 & $-$0.05 & $-$0.05 & 2280 \\
bm2-06 & 17:53:30.68 & -29:58:15.75 & 10.01 &  9.71 &  9.65 &  9.48   & 4208 & 	1.26 & 	$-$1.00   & 6.8  & $+$0.20  & $+$0.50 & 1200 \\
bm2-11 & 17:53:31.50 & -29:58:28.51 & 10.20 &  9.91 &  9.83 &  9.68   & 4005 & 	1.20 & 	$-$0.60	 & 7.2 & $+$0.25  & $+$0.30 & 1680 \\
bm2-12 & 17:53:31.74 & -29:58:22.94 & 10.67 & 10.32 & 10.30 & 10.09  & 	4003 &  1.67 & 	$+$0.15	 & 6.3 & $-$0.30 & $-$0.15 & 480  \\
bm2-13 & 17:53:31.14 & -29:57:32.76 & 10.86 & 10.52 & 10.50 & 10.29  & 	3727 & 	0.94 & 	$-$0.17	 & 6.9 & $+$0.13  & $+$0.40 & 2400 \\
bm2-15 & 17:53:30.23 & -29:56:42.74 & 10.43 &  9.96 & 10.07 &  9.74  & 3665 & 	1.09 & 	 $+$0.20 & 5.6 & 0.00  & $+$0.30 & 1200 \\
bm2-16 & 17:53:29.54 & -29:57:22.71 & 10.93 & 10.57 & 10.56 & 10.35  & 3886 & 	1.38 & 	 $+$0.06 & 3.14 & $-$0.02  & --- & 1200 \\
\hline
\hline
\end{tabular}
\end{table*}

\subsection{High spectral resolution VLT/CRIRES observations\label{ao}}

When possible, we used the  {\it Adaptive Optics} (AO) MACAO system when observing with the CRIRES spectrometer, concentrating the light into our slit.  
Special care was thus taken to find an optimal AO guide star, sufficiently bright  in the $R$ band (the band in which  the wavefront sensing is done, with $R<14$ required), and within $15\arcsec$ of our science target, in order for an optimal performance of the  MACAO system.

\begin{figure}
    \includegraphics[width=\columnwidth]{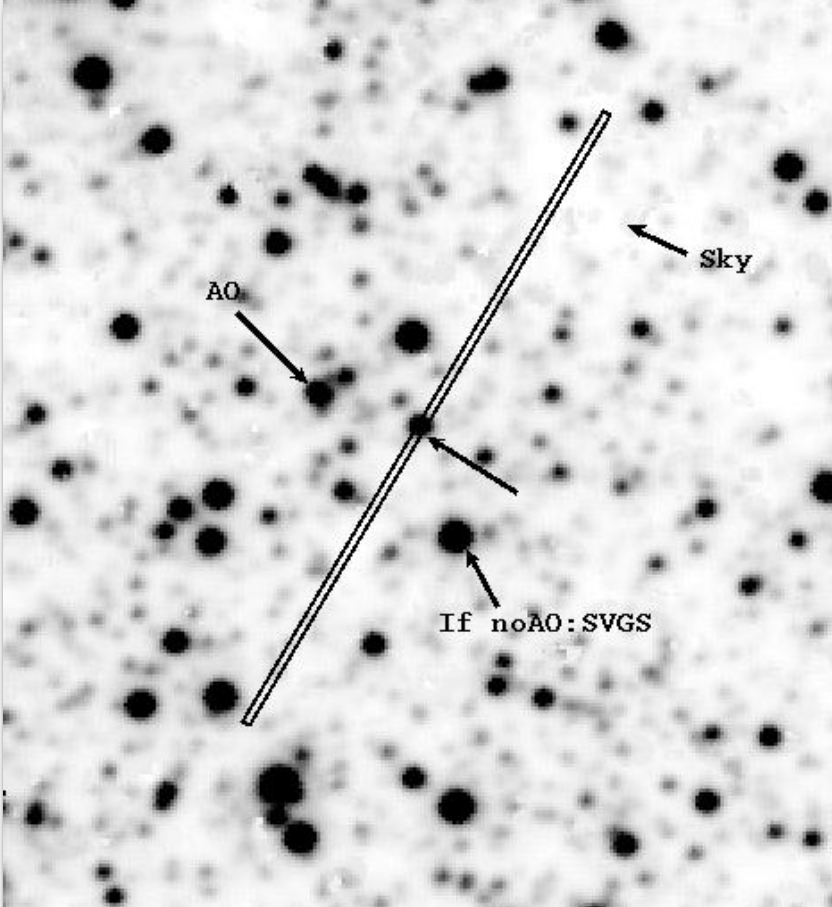}
    \caption{An example of a finding chart from our observing run with VLT/CRIRES. The underlying image is a DR9, UKIDSS K-band image \citep{ukidss}.   The observed giant star GC37, lies in the Galactic Center field. East is to the left and North stright down. The slit is $40\arcsec$.   }
    \label{fig:findingchart}
\end{figure}

Furthermore, special care was also taken to individually determine the position angles of the slit for every target star in these crowded fields.  This is done in order to ensure that we only record the target star on the $40\arcsec$ long slit, also when nodding along the slit and jittering to reduce the sky background, following standard procedures \citep{crires:manual}.  Figure \ref{fig:findingchart} shows an example of a finding chart in the K band, used at the observations, with the target and the AO-guide stars marked. The crowding of stars in these regions and the size of the long, narrow slit of the CRIRES spectrometer is illustrated nicely in the Figure and demonstrates the challenges in observing in the inner bulge. 

A slit width of $0.''4$, which we used, yields a resolving power of $R\sim 50,000$. For the CRIRES observations, we used a standard setting ($\lambda^\mathrm{ref}_\mathrm{vac}=2105.5$, order=27) with an unvignetted spectral range covering $20818-21444$\,\AA, with three gaps (20\,\AA) between the four detector arrays. Exposure times of $0.5 - 1$ hour per star provide a signal-to-noise ratios per pixel of typically 50-100. 

The reduction of all of the CRIRES observations was accomplished by following standard methods \citep{crires:cook} using Gasgano \citep{gasgano}. Subsequently, we used IRAF \citep{IRAF} to normalize the continuum, eliminate obvious cosmic hits, and correct for telluric lines (with telluric standard stars).

\subsection{Low spectral resolution VLT/ISAAC observations}
We obatined spectra using the red grism of the ISAAC spectrograph (ESO,VLT) covering
the wavelength range between $2.00-2.53\,$\mic. The slit width is 1\,$\arcsec$ providing a resolving power of $\rm R \sim 2000$ and the typical S/N ratio is about 100.

We observed B dwarfs (typically 6-8 stars per night) close to the airmass range of our targets, to use as telluric standard stars to correct for the instrumental and atmospheric transmission. We used IRAF to reduce the ISAAC spectra.
We removed cosmic ray events, subtracted the bias level, and then divided all frames by a normalized flat field. We used the traces of stars at two different positions (AB) along the slit to subtract the sky. After extracting and co-adding the spectra, we calibrated wavelengths using the Xe-lamp. The r.m.s of the wavelength calibration is better than 0.5\,\AA. The spectra were rebinned to a linear scale with a dispersion of  $\rm \sim 7 \AA $ and a wavelength range from 2.0\,$\mu$m to 2.51\,$\mu$m. Each spectrum has been divided by a telluric standard star observed closest in time and in airmass. 

\subsection{Low spectral resolution NTT/SOFI observations}  

We used the Red Grism of the SOFI spectrograph, covering 1.50--2.53 $\mu$m, to observe our M giant sample. We took the spectra with a 1\,$\arcsec$ slit providing a resolving power of  R $\sim$ 1000. We obtained  a $K_\mathrm{S}$-band acquisition image before each spectrum  to identify the source and place it on the 90\,$\arcsec$ slit. We used UKIDSS finding charts for source identification and to choose `empty' sky positions for optimal sky subtraction.

We observed B dwarfs (typically 6-8 stars per night), close to the airmass range of our targets, as telluric standard stars to correct for the instrumental and atmospheric transmission.  We used {\tt IRAF} to reduce the SOFI spectra. We removed cosmic ray events, subtracted the bias level, and then divided all frames by a normalized flat-field.  We used the traces of stars at two different (AB) positions along the slit to subtract the sky.  After extracting and co-adding the spectra, we calibrated wavelengths using the Xe-lamp. The r.m.s of the wavelength calibration is better than 1 \AA. We re-binned the spectra to a linear scale, with a dispersion of $\sim$ 10\,\AA /pixel.  We then divided each spectrum by the telluric standard observed closest in time and in airmass (airmass difference $<$ 0.05). We then normalized the resulting spectra by the mean flux between 2.27 and 2.29~$\mu$m.

\section{Analysis}

In order to determine the abundances from our spectra, we first have to determine the fundamental parameters of the stars. These are the effective temperatures (\teff), surface gravities (\logg), metallicities (\feh), and  microturbulences ($\xi_\mathrm{micro}$).
With these parameters, we can synthesize model spectra for the stars, based on spherical model atmospheres and an appropriate line list with atomic and molecular lines. In the following sections, we will describe the procedure to determine the stellar, fundamental parameters and the abundances of iron (metallicity), silicon, and magnesium from our spectra.

\subsection{Stellar Parameters}\label{params}

We determined stellar parameters in a similar way as for the Southern fields in \citet{ryde:16} using the relation between effective temperature and the
$\rm ^{12}CO$ band-head at 2.3\,$\mu$m. We refer for a more detailed discussion about this method in \citet{schultheis:16}. In addition we determined photometric effective temperatures from the dereddened $J_0$ and $K_0$ photometry.

The surface gravities were determined by the iterative method described in \citet{rich:17}, who demonstrated that this method determines surface gravities with a precision smaller than 0.3\, dex compared to an APOGEE data set. In short, we proceed as follows: we start off from a rough photometric estimate of the surface gravity by using $H$ and $K$ band photometry, extinction values (quite uncertain) from \citet{gonzalez2012}, and bolometric corrections from \citet{houdashelt2000}.
For this, we assumed a solar values of  $\rm T_{\odot}$=$5770\,K$, $\rm log g_{\odot}$=$4.44$, $\rm M^\mathrm{bol}_{\odot}$ =$4.75$, and a mass of our giants of 1.0\,$M_{\odot}$. With this surface gravity, a first metallicity estimate can be determined from our high-resolution CRIRES spectra as explained in the section\,\ref{abund}. A better estimate of the $\log g$ is subsequently found from the isochrone for a given \teff\ and [Fe/H].  We then use this new \logg\ to redetermine the \feh\ value. This procedure is iterated a few times until it converges.

The microturbulence, $\xi_\mathrm{micro}$, that takes into account the small-scale, non-thermal motions in the stellar atmospheres, is important for saturated lines, influencing their line strengths. We estimate this parameter from an empirical relation with the surface gravity based on a detailed analysis of spectra of five red giant stars (0.5 $<$ \logg $<$ 2.5) by \cite{smith:13}, as described in \citet{rich:17}. Our derived stellar parameters for all our stars are given in Tables \ref{starsall_N_tab} and \ref{starsall_S_tab}.

\subsection{K-Band Line List}

The line list used in the K band is based on an extraction from the VALD3 database \citep{vald,vald2,vald3,vald4,vald5}, but with updated values of wavelengths and line-strengths from recent experimentally and theoretically determined atomic line-strengths of Mg \citep{civis:13,pehlivan:mg}, as well as of Si (Pehlivan et al., 2018, in prep.). Furthermore, astrophysical $\log gf$-values are determined and wavelengths corrections are applied by fitting the solar center intensity atlas of \citet{solar_IR_atlas}. Of approximately 700
interesting spectral lines for cool stars identified in the K band, 570 lines have been assigned new values. 

The Fe, and Si abundances are determined from between 2-6 lines depending on star, telluric lines, radial velocities, and spectral quality. The Mg abundance is determined from two quite strong spectral features consisting of two fine-structure groups of 4f-7g lines at $21059-21061$\,\AA.
Since the pressure-broadening obviously affects this feature, we made calculations based on the Kaulakys method (\citet{kaulakys:85,kaulakys:91}), as described in \citet{osorio:15}. The line broadening cross section, $\sigma$, at a relative speed of $10^4$\,m\,s$^{-1}$ \citep[see][]{abo1,abo2,barklem:00} is found to be $2917\,$au [atomic units], expected to be accurate to 20-50\%, and the calculations indicate a roughly a flat temperature dependence, which was adopted in the calculations.

In the abundance analysis, we also include molecular line lists of the CN molecule \citep{sneden:14}.  

\subsection{Stellar Abundances} \label{abund}

\begin{figure}
    \includegraphics[width=\columnwidth]{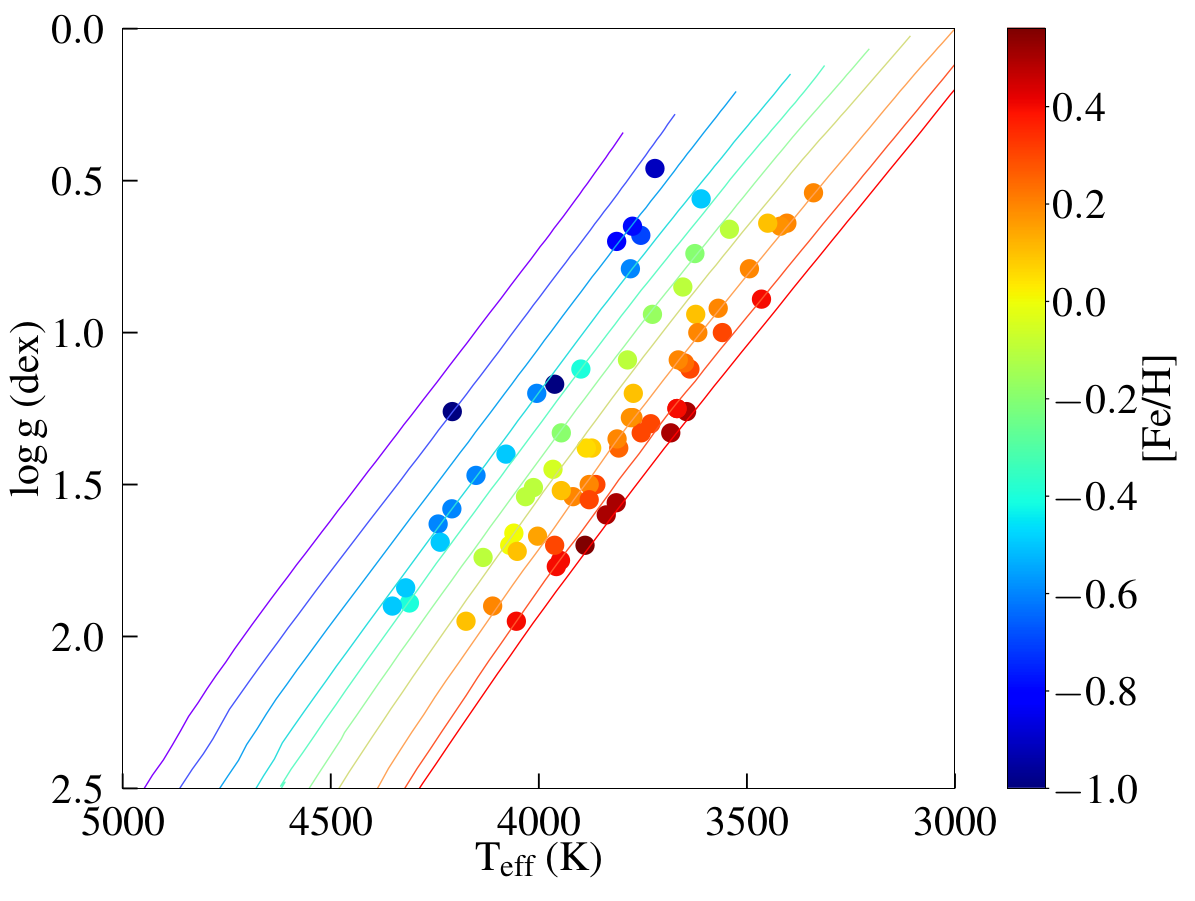}
    \caption{\teff\, vs \logg\, diagram showing the position of our sample of stars on different metallicity tracks of the 10 Gyr YY isochrone \citep{demarque:04}. Each star is color coded with their determined \feh\, values.  }
    \label{fig:HR}
\end{figure}
 
As mentioned above, once we have determined the \teff, an approximate \feh, initial photometric \logg-estimate, and the corresponding $\xi_\mathrm{micro}$ for our stars, we made use of our high resolution spectra to iteratively determine their final metallicities. Subsequently we can then determine the $\alpha$-element abundances (Si and Mg), see also \citet{rich:17}. 

We have used the package Spectroscopy Made Easy (SME) (\citealt{sme,sme_code,SME:2017}) for the spectral synthesis and the abundance analysis. SME generates synthetic spectra for a given set of fundamental, stellar parameters by interpolating within a grid of MARCS spherical-symmetric, LTE model atmospheres \citep{marcs:08}. The spectral lines of interest are marked with masks to indicate the parts of the lines that should be modeled and compared with the observed spectra. In a similar fashion, we also marked regions, around the lines being analyzed with masks, that should be treated by the SME as continuum regions. These are used for a final, local continuum normalization. Abundances or any other parameters to be determined are set as free variables, for which SME synthesizes spectra in an iterative manner following an approach that minimizes the $\chi^{2}$ by comparing the synthesized with the observed spectrum. 

Since SME compares the strengths and forms of observed spectral lines with the synthesized ones (within the defined masks), the line profiles have to be accurately characterized and the line broadening well determined. Extra broadening not accounted for by the synthesis, is the {\it stellar} macroturbulence $\xi^\mathrm{stellar}_\mathrm{macro}$, but also the instrumental profile, which is set by the instrumental resolution. We estimate this extra {\it total} broadening (which is a convolution of the both) for each of our stars by fitting a few selected well-formed, medium-weak lines while setting `$\xi_\mathrm{macro}$' as a free parameter in a SME run and assuming both broadening profiles to be Gaussian in shape. We then get the total macroturbulences for our stars, $\xi_\mathrm{macro}$, which are given in Tables \ref{starsall_N_tab} and \ref{starsall_S_tab}.

Subsequently, we ran SME for each star with the above mentioned parameters, letting SME perform a $\chi^{2}$ minimization to fit Fe lines in the spectra that are of sufficient quality, in shape and form (due to their strengths for a given SNR), for retrieving an abundance. We then adjusted the \feh\ value based on the fit, simultaneously changing to a new \logg-estimate based on position on the isochrone tracks, and a corresponding $\xi_\mathrm{micro}$ from the \cite{smith:13} relation. This is repeated until the stellar parameters are consistent with the corresponding isochrone in the HR diagram. Figure~\ref{fig:HR} shows the HR diagram with the locations of our stars on the 10 Gyr YY isochrones \citep{demarque:04}, color coded with their determined metallicities. With the fundamental stellar parameters thus determined, we proceeded to determine the [Si/Fe] and [Mg/Fe] using SME with similarly selected Si and Mg lines.

The final metallicities and Si and Mg abundances are given in Tables \ref{starsall_N_tab} and \ref{starsall_S_tab}.

\begin{table}
\caption{Uncertainties in the derived metallicities and magnesium abundances due to uncertainties in \teff\, of $\pm$\,150\,K, corresponding change in \logg\, and $\xi_\mathrm{micro}$. This was estimated using synthetic spectra with metallicities in the range of -0.8, -0.5, -0.3, 0.0 ,0.3 and 0.5 dex each with \teff\ of 3500, 3900 and 4300\,K, that represent the stars in our sample.  }\label{uncertainties_tab}
\begin{tabular}{l c c c c c c c}
 [Fe/H] & \teff  & $\delta$\,\teff & $\delta$\,\logg & $\delta$\,$\xi_\mathrm{micro}$& $\delta$\,[Fe/H] & $\delta$\,[Mg/Fe]\\
\hline
\hline
\multirow{4}{*}{-0.8} & \multirow{2}{*}{3900} & -150 &  -0.27 & +0.1 & -0.07 & -0.04\\
 &  & +150 & +0.27 & -0.2 & +0.19 & +0.10 \\
\\
 & \multirow{2}{*}{4300} & -150 & -0.27 & 0.0 & +0.01 & +0.07\\
 &  & +150 & +0.28 & -0.1 & +0.09 & +0.06\\
\hline
 \multirow{6}{*}{-0.5} & \multirow{2}{*}{3500} & -150 & -0.25 & +0.1 & -0.06 & +0.03\\
 & & +150 & +0.25 & -0.2 & +0.03 & -0.01\\
\\
 &\multirow{2}{*}{3900} & -150 &  -0.27 & +0.1 & -0.09 & -0.04\\
 &  & +150 & +0.28 & -0.2 & +0.07 & +0.11\\
\\
 & \multirow{2}{*}{4300} & -150 & -0.28 & +0.1 & -0.07 & -0.03\\
 &  & +150 & +0.29 & 0.0 & +0.06 & +0.01\\
\hline
 \multirow{6}{*}{-0.3} & \multirow{2}{*}{3500} & -150 & -0.24 & +0.1 & -0.10 & 0.00 \\
 & & +150 & +0.25 & -0.2 & -0.02 & -0.06\\
\\
 &\multirow{2}{*}{3900} & -150 &  -0.27 & +0.1 & -0.14 & -0.11\\
 &  & +150 & +0.28 & -0.1 & -0.03  & -0.06\\
\\
 & \multirow{2}{*}{4300} & -150 & -0.28 & +0.1 & -0.09 & -0.04\\
 &  & +150 & +0.29 & 0.0 & +0.05 & -0.02\\
\hline
 \multirow{6}{*}{0.0} & \multirow{2}{*}{3500} & -150 & -0.24 & +0.1 & +0.02 & +0.10 \\
 & & +150 & +0.26 & -0.2 & +0.05 & -0.01\\
\\
 &\multirow{2}{*}{3900} & -150 &  -0.28 & +0.1 & -0.01 & -0.05\\
 &  & +150 & +0.28 & 0.0 & +0.02 & -0.17\\
\\
 & \multirow{2}{*}{4300} & -150 & -0.30 & 0.0 & -0.06 & +0.04\\
 &  & +150 & +0.31 & -0.1 & +0.07 & 0.00\\
\hline
 \multirow{6}{*}{0.3} & \multirow{2}{*}{3500} & -150 & -0.25 & -0.1 & 0.00 & +0.06 \\
 & & +150 & +0.26 & -0.2 & +0.17 & +0.08\\
\\
 &\multirow{2}{*}{3900} & -150 &  -0.27 & 0.0 & -0.01 & +0.18\\
 &  & +150 & +0.28 & -0.1 & +0.04 & -0.05\\
\\
 & \multirow{2}{*}{4300} & -150 & -0.31 & +0.1 & -0.06 & +0.06\\
 &  & +150 & +0.36 & 0.0 & +0.08 & -0.06\\
\hline
 \multirow{6}{*}{0.5} & \multirow{2}{*}{3500} & -150 & -0.26 & +0.1 & -0.05 & +0.06 \\
 & & +150 & +0.26 & -0.1 & +0.06 & +0.11\\
\\
 &\multirow{2}{*}{3900} & -150 &  -0.27 & +0.1 & -0.04 & +0.09\\
 &  & +150 & +0.27 & 0.0 & +0.06 & -0.07\\
\\
 & \multirow{2}{*}{4300} & -150 & -0.32 & +0.1 & -0.07 & +0.09\\
 &  & +150 & +0.37 & 0.0 & +0.08 & -0.07\\
\hline
\hline
\end{tabular}
\end{table}

\begin{figure*}
    \includegraphics[width=\columnwidth]{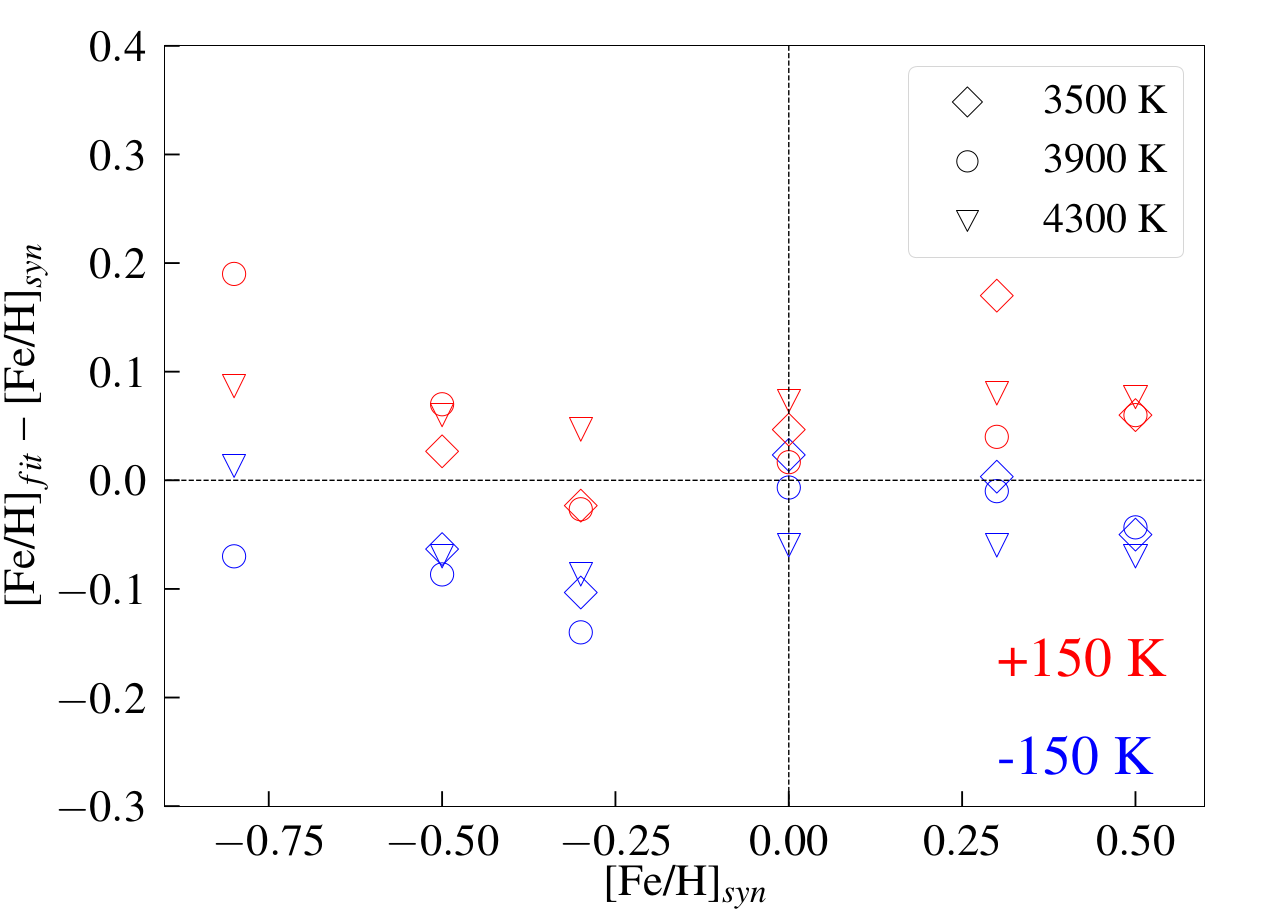}
    \includegraphics[width=\columnwidth]{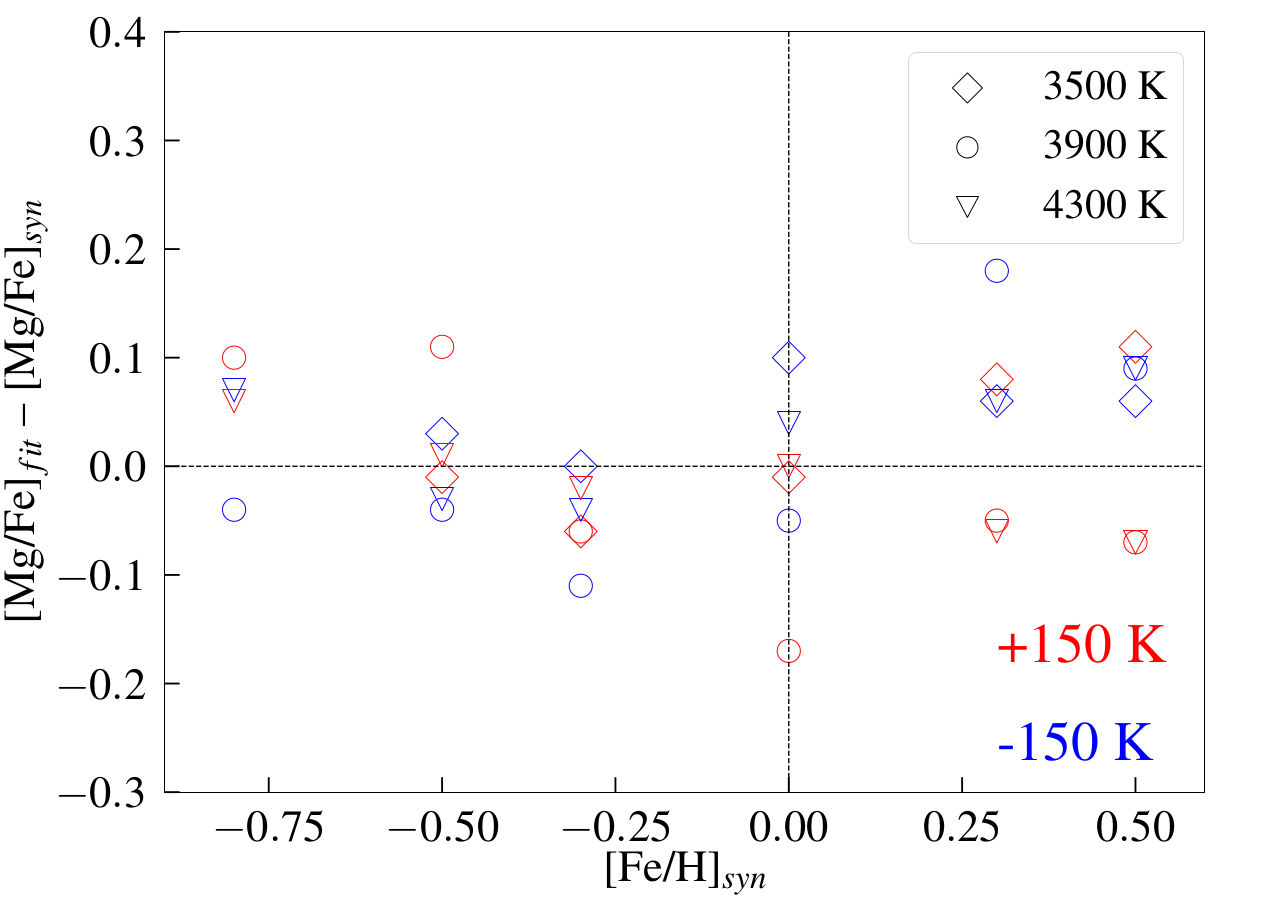}
    \caption{Plots showing the trend of \feh\, and \mgfe\, uncertainties as a function of metallicity for the case of different temperatures covered by our sample. Left panel shows the differences in SME estimated metallicities vs actual metallicities due to uncertainties in \teff\, of $\pm\,150$\,K, corresponding change in \logg\, and $\xi_\mathrm{micro}$. Right panel shows the same for magnesium abundances. This was estimated using synthetic spectra with metallicities in the range of $-0.8$, $-0.5$, $-0.3$, $0.0$,$0.3$ and $0.5$ dex each with \teff\ of 3500, 3900 and 4300\,K, that represent the stars in our sample.}
    \label{fig:alpha_fe_uncertainty}
\end{figure*}

\subsection{Abundance Uncertainties}

It has been shown in \citet{rich:17,ryde:16} that the uncertainties in the derived metallicities arise partly from the fitting procedure and partly from the uncertainties in the derived stellar parameters like \teff, \logg, and $\xi_\mathrm{micro}$. The systematic uncertainties due to the noise in the spectra, telluric residuals and the uncertainty in continuum level  is assumed to be of the order of 0.15\,dex. When observing our stars we aimed for a S/N of 60-70 per pixel, which means that the uncertainties are of similar magnitudes for all the stars. For a few stars (e.g. GC25 and GC44) the S/N is lower, and these uncertainties can therefore be larger.  

Even though the above mentioned studies have explored the uncertainty in metallicity due to a variation in stellar parameters for individual stars of a given \teff\ and metallcity, the variation of the uncertainty in metallicity determination for a range of metallicities and \teff\ has been relatively unstudied. We have carried out such an investigation in Section \ref{uncert}.

\subsubsection{General Uncertainties\label{uncertgeneral}}

The abundance is given by the line strength, which is the contrast between the continuum and the line center. Not only the determination of the continuum level in the observations, which can be difficult depending on SNR and spurious feature in the spectrum, is crucial, but also the calculation of the continuous opacity in the generation of the synthetic spectra is important. For a typical star in our sample, the continuous opacity is due to H$^{-}$ free-free opacity, which is affected by the electron density. This is given by the major electron donors in the line-forming regions of the star, which are magnesium (1/2 of all electrons), iron (1/3 of electrons), and Si (1/10). Thus, it is especially important to know the magnesium abundance for an accurate abundance determination of any element. We have taken the general $\alpha$-element trend vs. metallicity for bulge stars into account in order to minimize this effect. Also, special care has been taken to find good continuum points. 

As described above, the way SME works by fitting the shape of the line, the determined abundance is sensitive to the line broadening adopted, most importantly the macroturbulence but also the pressure broadening for strong lines, like the Mg lines used in this study. We have therefore invested considerable effort to estimate the broadening as accurately as possible. 

Most line strengths in our line list, are determined from the solar spectrum (astrophysical $gf$-values). We have also tested the list against the spectra of the metal-poor, $\sim$4800\,K giant $\alpha$ Boo, and the metal-rich giant $\mu$ Leo. There might, however, be non-LTE, saturation, or 3D effects that are not taken into account which might affect different type of stars differently, see \citet{rich:17} for a discussion. We have avoided spectral lines that we suspect are affected by these issues.

As mentioned above, we estimate the total uncertainty due to the fitting procedure to approximately 0.15 dex.

\subsubsection{Uncertainties related to stellar-parameters \label{uncert}}

We synthesized a grid of test spectra of similar spectral resolution as that of our observed spectra, with metallicities of -0.8, -0.5, -0.3, 0.0, 0.3, 0.5\,dex each with \teff\, of 3500, 3900, 4300\,K.  For each of these test spectra, we determined the surface gravities and the corresponding microturbulence in the same way as for our observed spectra (see Section \ref{params}).

For each  of the test spectra, we varied \teff\, by the typical uncertainty from the low-resolution, CO-bandhead estimate, namely $\pm$150\,K, changed \logg\, as well as $\xi_\mathrm{micro}$ correspondingly, and then ran SME with these parameters, setting \feh\, as free parameter. This exercise was carried out for each of the three \teff\, cases for all six \feh\ values, thus covering the full range of fundamental stellar parameters of our observed sample. The difference between the actual \feh\, and that estimated by SME gives us the typical uncertainty in \feh\, that arises from the combination of errors in \teff, \logg, and $\xi_\mathrm{micro}$. The same exercise was carried out to estimate the typical uncertainty in \mgfe\, by setting Mg as the free parameter. Table~\ref{uncertainties_tab} lists the final estimated error in \feh\, and \mgfe\,, and we plot them in  Figure~\ref{fig:alpha_fe_uncertainty}. The typical change in \logg\, ranges from $\pm$0.25 to 0.37\,dex and in $\xi_\mathrm{micro}$ from $-0.2$ to $+0.2$\,km s$^{-1}$, for a $\pm$150\,K change in \teff. The maximum values of \feh\, uncertainties are of the order of $0.2$\,dex, found partly at the lowest metallicity and partly at low \teff, whereas the maximum uncertainty of \mgfe\ of the same order, is found at supersolar metallicities.



Indeed, one could suspect that there should be a systematic trend in the uncertainties due to the line strength. Lines tend to get weaker for more metal-poor stars (\teff\ and \logg\ may, however, also play a role). Thus, for a given SNR, metal-poor stars might have weaker lines, and therefore more uncertainty in the derived abundances. 

Furthermore, the cooler the star, the stronger the molecular features are, and therefore the quality of the molecular line list is increasingly important. In addition,  the continuum determination might be more uncertain. In no case were the molecular lines so dense as to prevent us from identifying continuum points. Hence we do not require pseudo-continua, i.e. a situation where the true continuum is not found but many weak lines blend into each other forming flat regions, resembling a continuum. 

Also, for more metal-rich stars, there might be a problems with saturation of the lines, especially for the strong Mg lines, which get increasingly saturated and therefore less abundance sensitive.

However, from Table~\ref{uncertainties_tab} and Figure~\ref{fig:alpha_fe_uncertainty} we see that the uncertainties do {\it not dramatically} increase neither towards the metal-poor regime, nor the metal-rich one, which is reassuring. Nevertheless we find that warmer stars have generally lower abundances uncertainties, less than 0.1\,dex. 


We conclude that apart from more systematic uncertainties (e.g. continuum placement) the typical abundance uncertainties are due to errors in the stellar parameters, which are $\sim$<0.1\,dex.  We have somewhat higher errors for stars with [Fe/H]<-0.5\,dex, and further, the Mg abundances for the coolest and most metal rich stars are uncertain by $\sim$0.15\,dex.

A grand total uncertainty is thus mainly due systematic fitting errors (Section \ref{uncertgeneral}), which leaves us with a total estimate of approximately $0.15$\,dex.

\begin{figure*}
    \includegraphics[width=\textwidth]{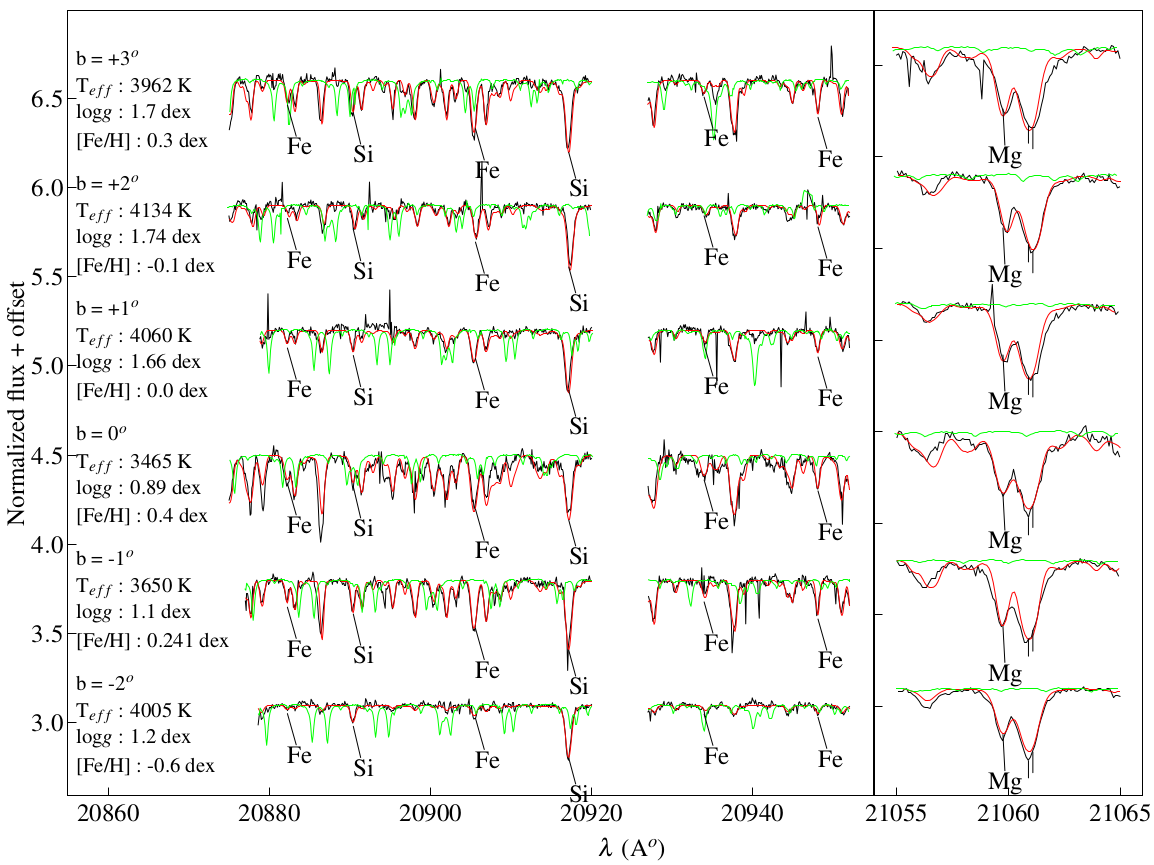}
    \caption{Spectra of wavelength regions covering a few Fe lines used for the abundance determination. The latitudes and their stellar parameters of each star are indicated beside each spectrum. Synthetic spectra are shown in red and the telluric lines, shown in green, indicate the areas where they hit the spectrum. Also indicated are two Si lines used in the determination of Si abundance. Rest of the features seen in the spectra are mostly CN molecular lines. This shows how well they are separated from our lines of interest (Fe and Si) indicating the advantages of high resolution spectra and the quality of molecular CN line-lists from \protect\cite{sneden:14}.}
    \label{fig:ns_spectra}
\end{figure*}

 
 
 
 


\subsection{Homogeneous Analysis of the Entire Sample}

The 44 giants in the Northern fields are analysed here for the first time. The 28 stars in the Southern fields  were first published in \citet{ryde_schultheis:15,ryde:16}, but are reanalysed here in the same manner as the Northern-field giants, to allow for the whole sample to be analysed as consistently as possible. This will minimize the relative uncertainties and will allow for a comparison between properties of the stars in the different fields. 

Specifically, the 9 stars in the Galactic center field were first published in \citet{ryde_schultheis:15}, and the  9 stars at ($l,b$) = ($0,-1^{\circ}$), and 10 stars at  ($l,b$) = ($0,-2^{\circ}$) were published in \citet{ryde:16}. The largest difference between the analyses in these publications and our analysis, is the method we use to determine the surface gravities of the stars. Here, we force the stars to be consistent with isochrones, following  the method outlined and described in \citet{rich:17}, instead of the more uncertain method of assuming the stars to be at a certain distance (8 kpc) and dereddening the photometry. Differences as large as $\Delta \log g$ =$0.6$ can be found. As described in \citet{schultheis:17} the latter method can give large uncertainties, since the Galactic bulge shows an intrinsic depth ($\rm \sim 1-2\,kpc$)  which can result in large uncertainties in log\,g, especially for M giants. We refer here for  a more detailed discussion to \citet{schultheis:17}.
The new surface-gravity determinations can account for most of the changes in the derived abundances.  Furthermore, for a few of the stars a different value of the macroturbulence is determined. Due to the way the $\chi^2$ minimization is done in line masks, the derived abundances are quite sensitive to the broadening fit of the lines. For yet another few stars, most notably GC25 and GC44, the abundance sensitive lines in the spectra are affected by so much noise that the determination is quite uncertain. This was not the case for most of our sample, however.




Thus, we are confident that we can make a comparative study of the stellar properties and the metallicity distributions in the different location along the minor axes, i.e. metallicity gradients, and the symmetries of the Northern fields compared to the Southern fields.

\section{Results} \label{result}

\begin{figure}
    \includegraphics[width=\columnwidth]{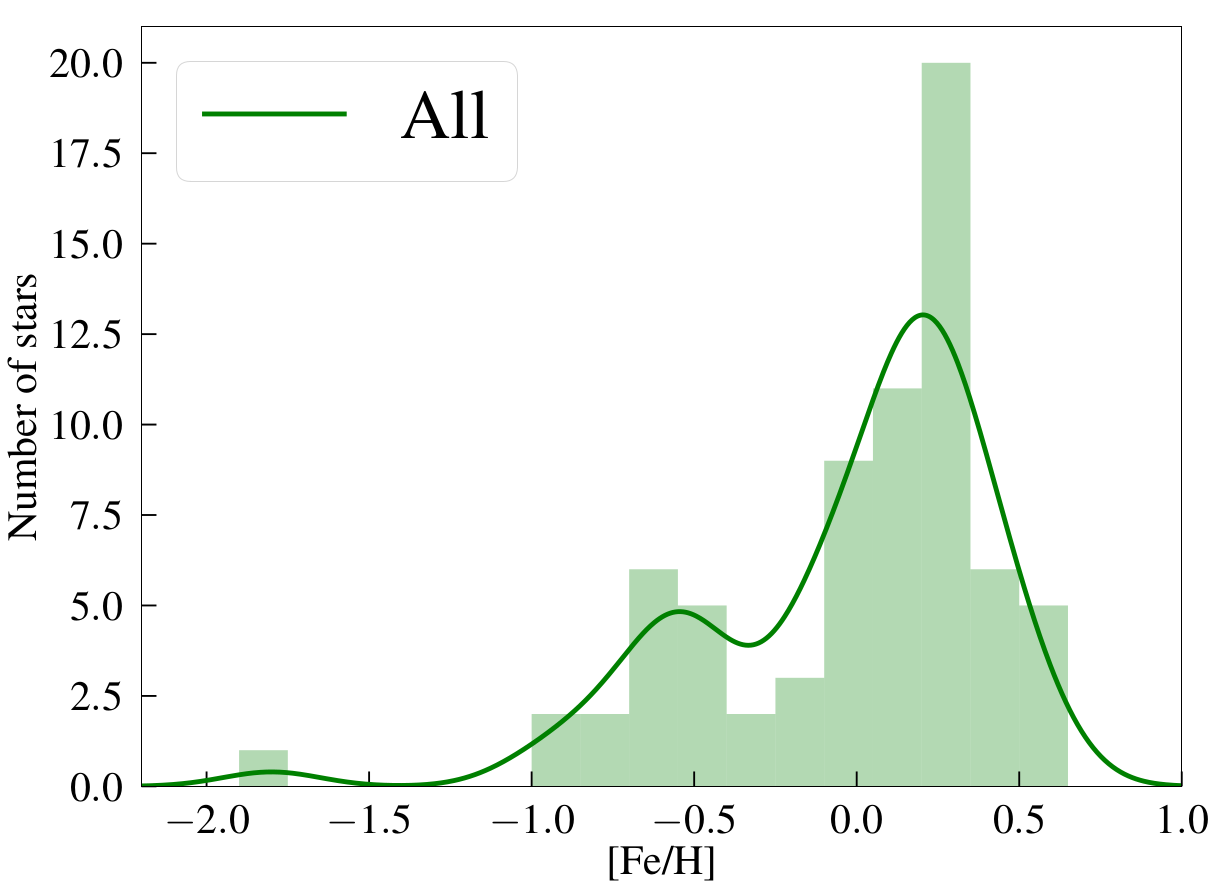}
    \caption{MDF for our entire sample in bins of 0.15\,dex, overlaid with kernel density estimate (KDE) with bandwidth of 0.5\,dex (green line).}
    \label{fig:ns_all}
\end{figure}

\begin{figure}
    \includegraphics[width=\columnwidth]{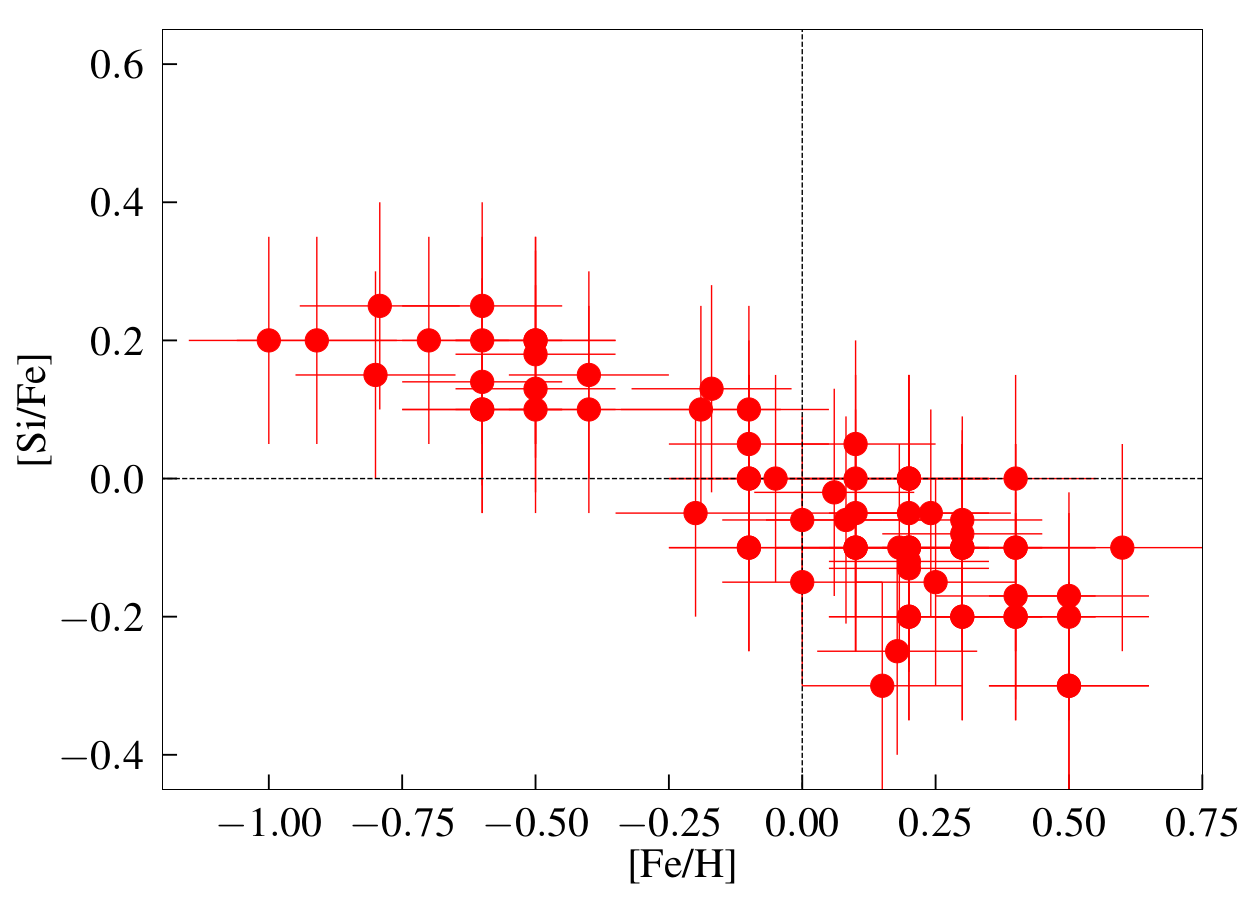}
    \includegraphics[width=\columnwidth]{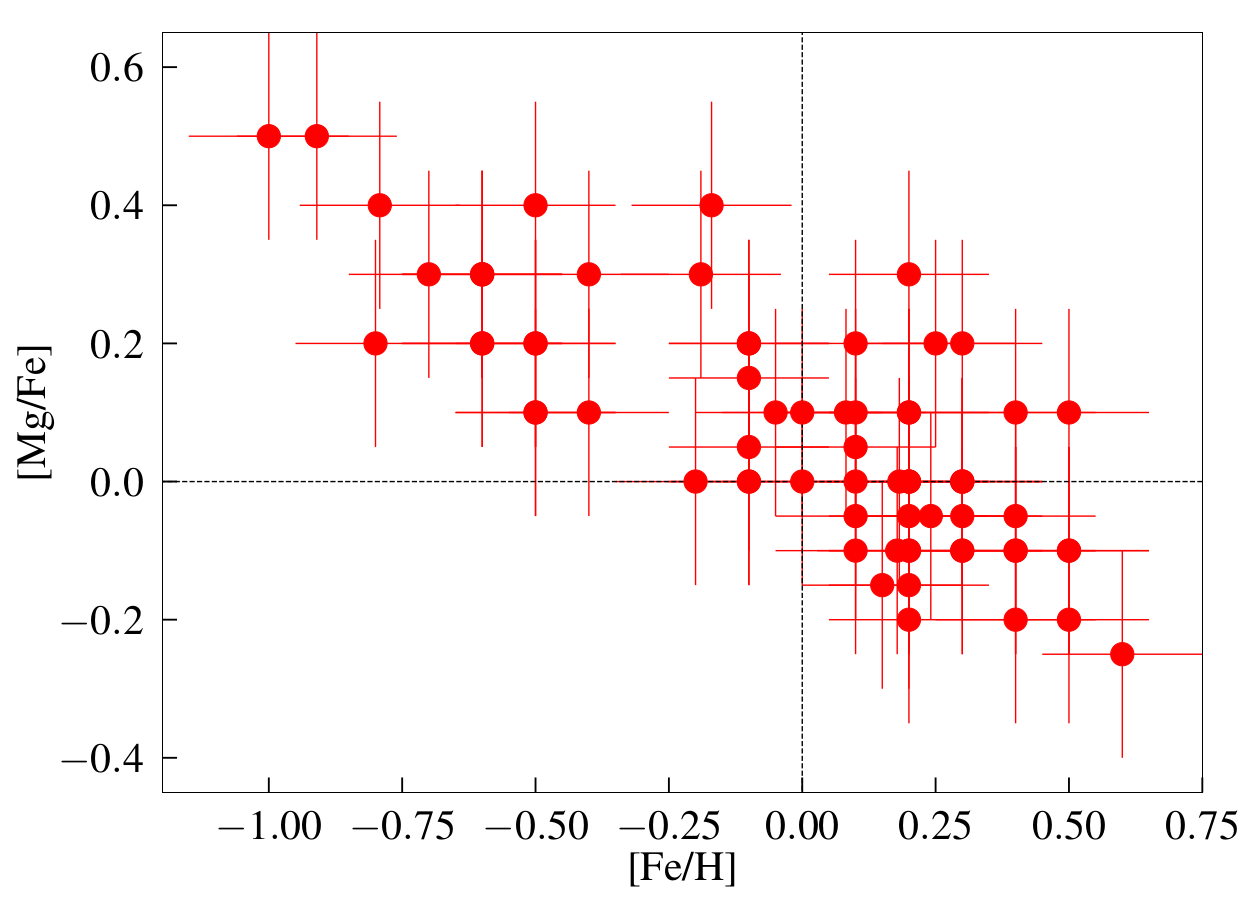}
    \caption{The \sife\, (top panel) and \mgfe\, (bottom panel) trends as a function of the metallicity for the entire sample, with the typical uncertainty of 0.15\,dex}
    \label{fig:alpha_fe}
\end{figure}

\begin{figure}
    \includegraphics[width=\columnwidth]{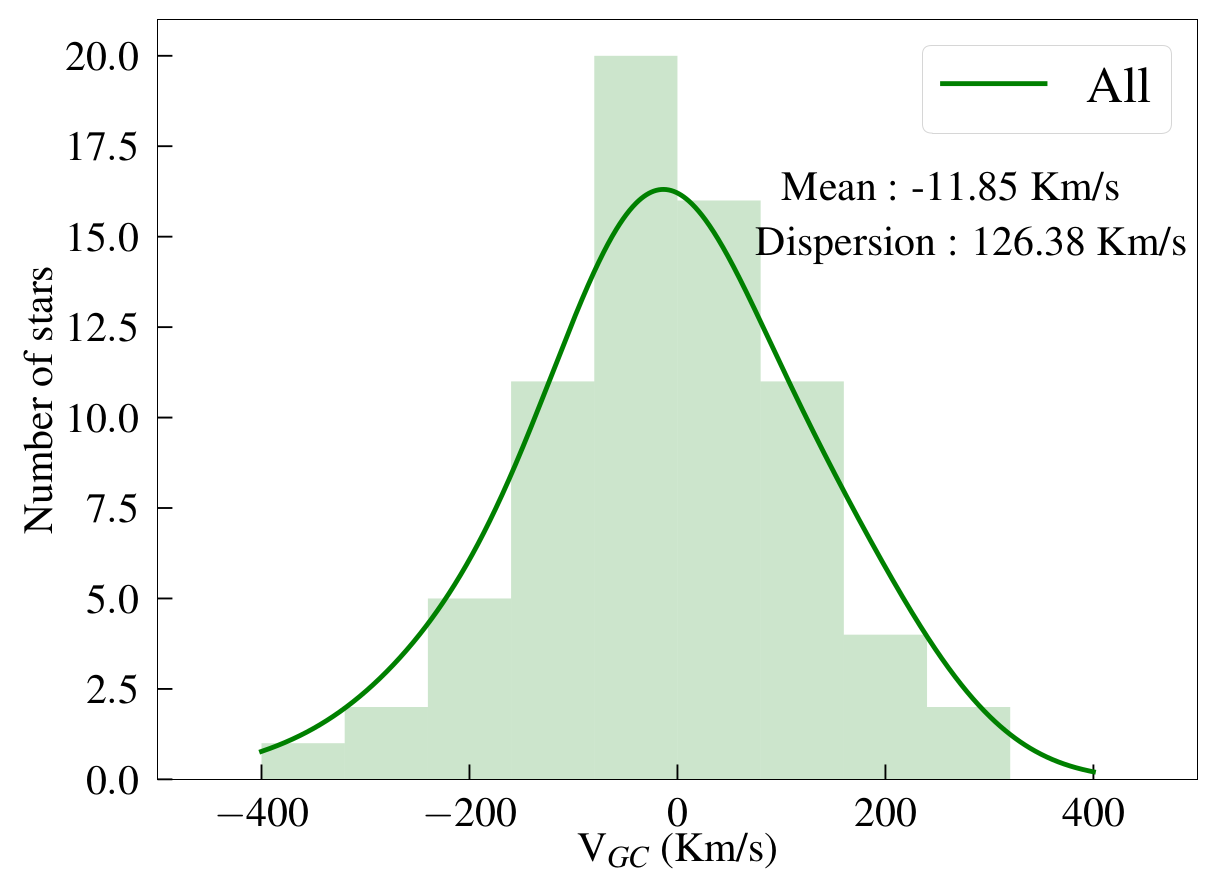}
    \caption{Galacotcnetric radial velocity distribution for our entire sample in bins of 80 kms$^{-1}$, overlaid with KDE with the same bandwidth. The mean and dispersion estimated for the sample are indicated in the plot.}
    \label{fig:vrad_all}
\end{figure}

The final reduced spectra and fitted synthetic spectra for one star in each field are shown in Fig~\ref{fig:ns_spectra}. The location of several Fe, Si and Mg lines used for our abundance estimation are also shown in the plots. The necessity of requiring high resolution spectra for abundance analysis can be evidently seen, as we are able to clearly separate the molecular CN lines in our spectra. These features can lead to more blending in Fe and Si lines in lower resolution spectra. The stellar parameters for all stars belonging to the Northern and Southern fields along with their derived $\alpha$-element abundances of \sife\, and \mgfe\, are listed in Tables~\ref{starsall_N_tab} and ~\ref{starsall_S_tab}. They are normalized to the solar abundances of \cite{solar:sme}. 

We plot the MDF of our entire sample in bins of 0.15\,dex as shown in Figure~\ref{fig:ns_all} together with  the kernel density estimate (KDE) of the MDF  with a bandwidth of 0.15\,dex.  We see clearly a bimodal  distribution, one at metallicity of $\sim$+ 0.3\,dex and the other at $\sim$-0.5\,dex. An outlier with metallicity of -1.8\,dex leads the KDE estimate to find a small peak at this metallicity. Keeping this star aside, the metallicity of our entire inner bulge sample along the bulge minor axis is in the range of $-1.0<$\feh\,$<0.6$\,dex.

In Figure~\ref{fig:alpha_fe}, we show the \sife\, and \mgfe\, trends as a function of the metallicity for the entire sample, with the typical uncertainty of 0.15\,dex. We find the expected trend of supersolar $\alpha$-abundances for metal-poor stars, while metal-rich stars show subsolar $\alpha$-abundances. Though the \sife\, abundances follow this trend, they are systematically lower than \mgfe\, values in the subsolar metallicity range. The \sife\, abundances  show a lower dispersion than \mgfe\, abundances which is most likely due to the fact that we use several Si lines for the Si-abundance while for Mg we use only one line feature. We see clearly
that for supersolar metallicities the Si and Mg abundances  decrease with increasing
metallicity.

In addition, we measured the approximate heliocentric radial velocity, V$_{R}$ of each star by estimating the difference in the wavelength position of a strong line in their observed spectra with that from the laboratory measurement (Mg line at 21061.095\,nm). Based on spectral resolution, one pixel in velocity space corresponds to $\sim$6\,kms$^{-1}$, which can be considered to be the typical uncertainty in V$_{R}$. We converted this to Galactocentric radial velocity, V$_{GC}$, by adopting the local standard of rest velocity at the Sun to be 220\,kms$^{-1}$ and a solar peculiar velocity of 16.5\,kms$^{-1}$ in the direction (l,b) = (53\degr,5\degr) \citep{ness:13a}. 

\begin{equation}
	$$V_{GC}$=$V_{R}$+$20\,$[$\sin$($l$)$\cos$($b$)$$]+$16.5$[$\sin$($b$)$\sin$($25$)+$\cos$($b$)$\cos$($25$)$\cos$($l-53$)]$$$
\end{equation}

We show the Galactocentric radial velocity distribution for our entire sample in  Figure~\ref{fig:vrad_all}. We find a mean value of $-12$\,kms$^{-1}$ with a dispersion of $126$\,kms$^{-1}$.



\section{Discussion}

Here, we will discuss and compare our results with other literature studies carried out in the inner and outer Galactic bulge fields. We also investigate for the first time, the symmetry in MDF and $\alpha$ abundance trends between the Northern and Southern fields of inner Galactic bulge.






\begin{figure*}
    \includegraphics[width=\columnwidth]{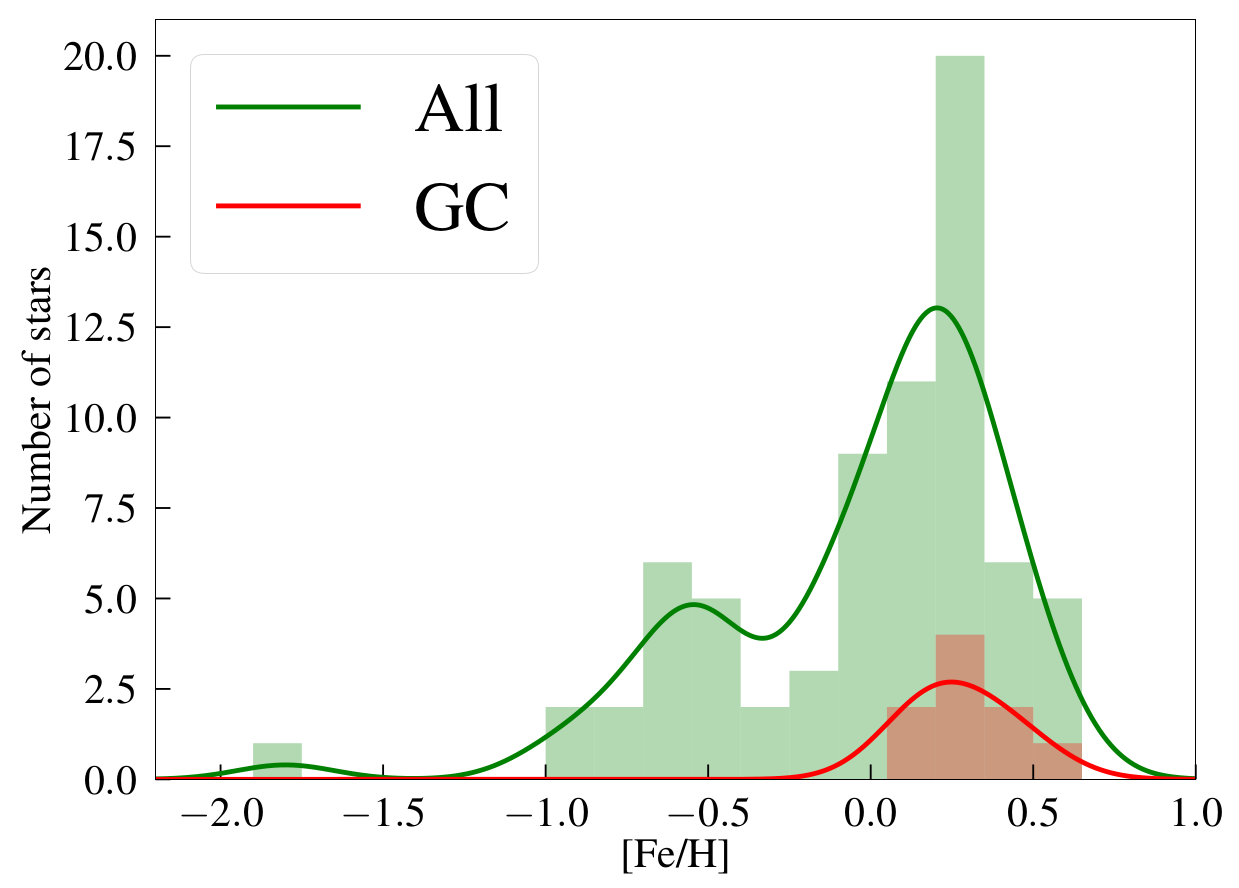}
     \includegraphics[width=\columnwidth]{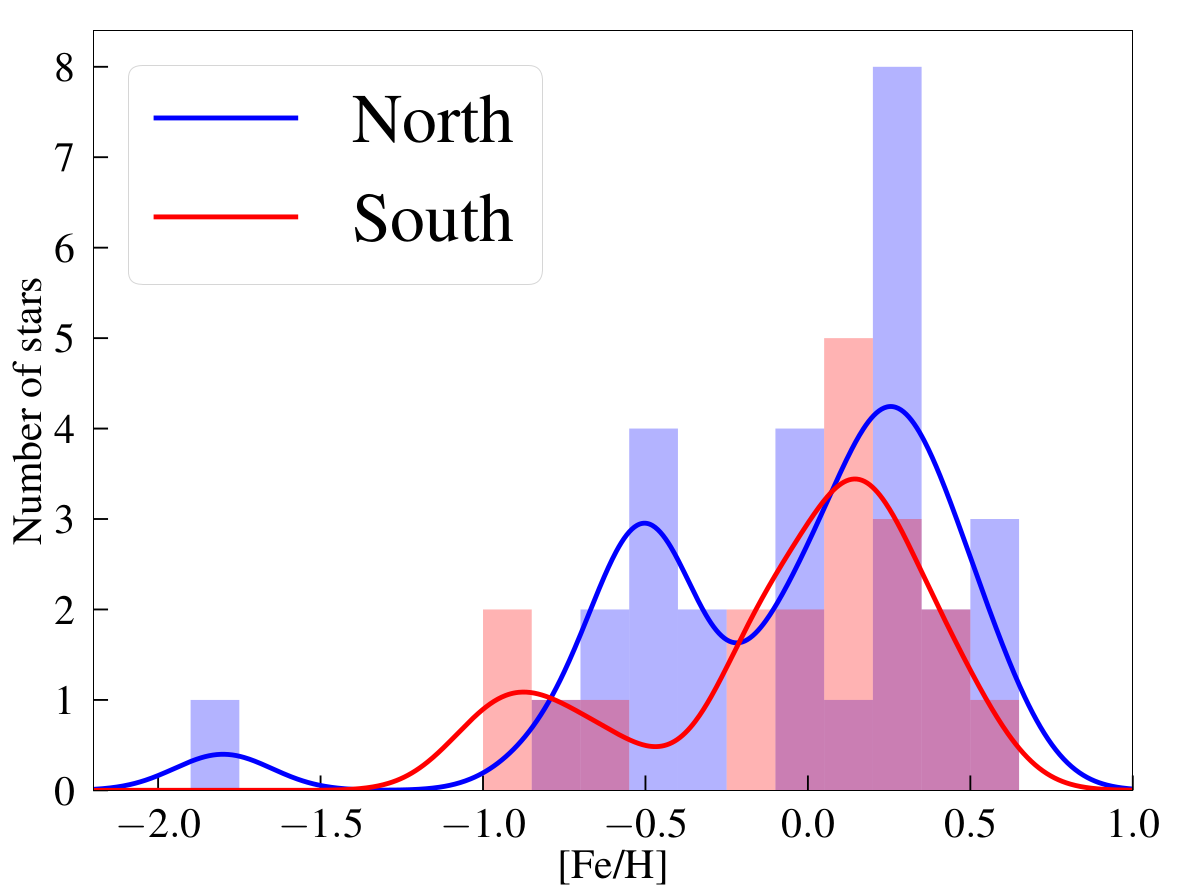}
    \caption{(left) MDF of the Galactic center sample (red) in comparsion with that of the entire sample (green). Galactic center stars in our sample are all metal rich. (Right) MDF of North (blue; b = +1\degr, +2\degr) and South (red; b = -1\degr, -2\degr) field stars to investigate the North-South symmetry in MDFs. Overlaid KDEs,shown in respective colors, use the same bandwidth as binsizes of the histograms.}
    \label{fig:ns_all_vs_gc}
\end{figure*}

\begin{figure*}
    \includegraphics[width=\columnwidth]{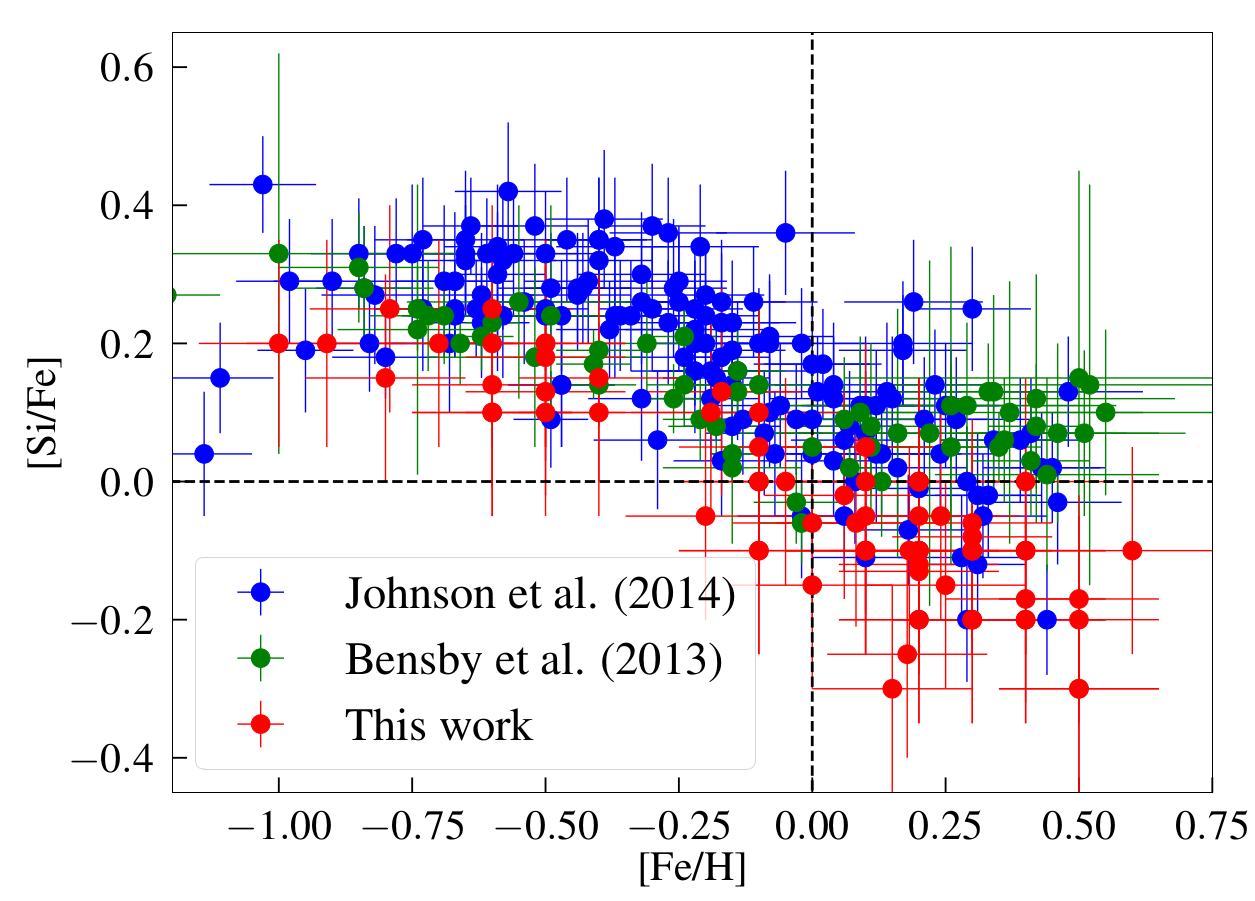}
    \includegraphics[width=\columnwidth]{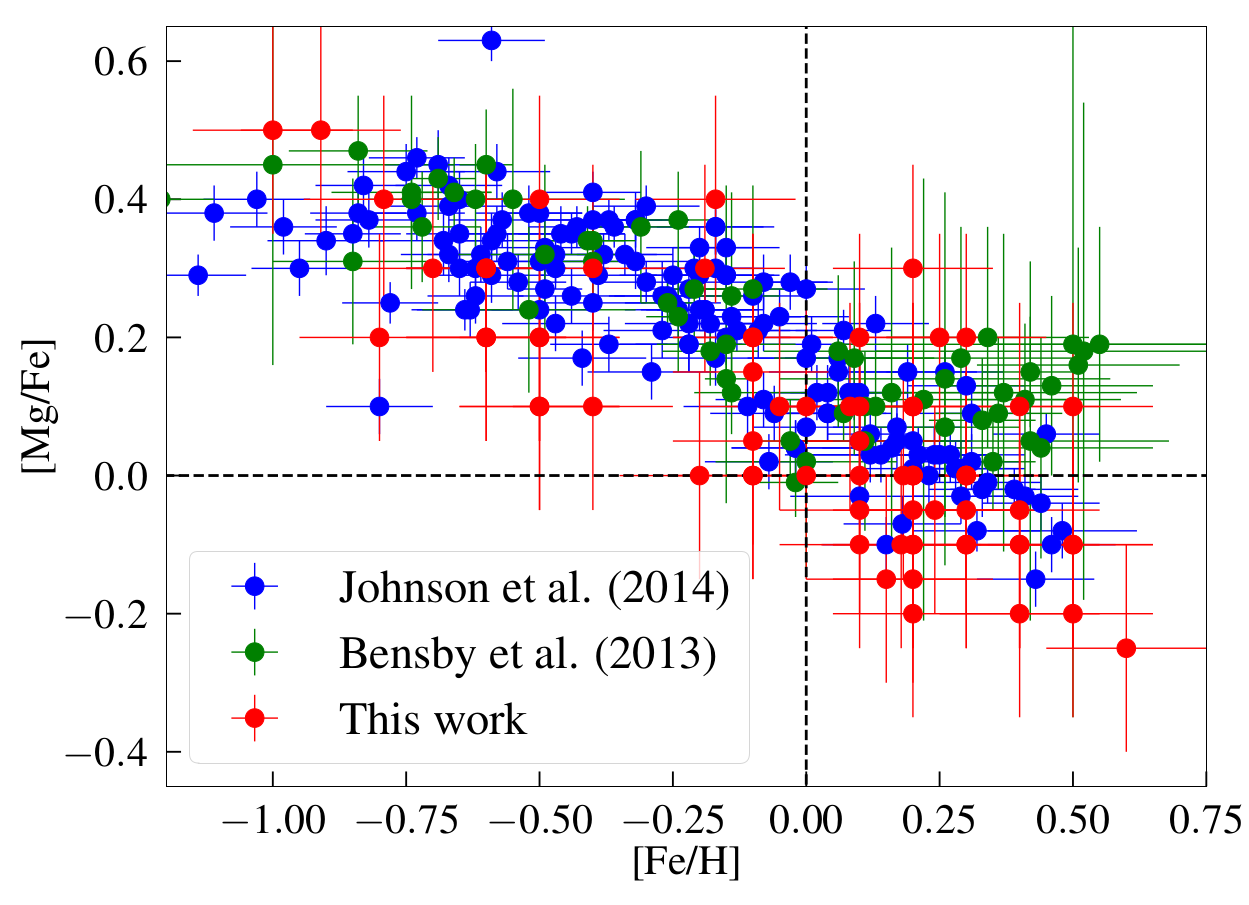}
    \caption{[Si/Fe] vs [Fe/H] (left) and [Mg/Fe] vs [Fe/H] (right) trends of our sample (red filled circles) in comparison with that of the micro-lensed dwarf sample in \protect\cite{bensby:13} (green filled circles) and red giant sample in \protect\cite{johnson:2014} (blue filled circles) in outer bulge fields. We plot our sample with typical uncertainty of 0.15\,dex, while the comparison samples are plotted with individual uncertainties of each star estimated in \protect\cite{bensby:13} and \protect\cite{johnson:2014}. Our trends in the subsolar metallicity regime are consistent within the 1-$\sigma$ uncertainty limit of both samples. In the supersolar metallicity regime, we find subsolar alpha-abundances causing the trend to go down, consistent with \protect\cite{johnson:2014}. Meanwhile the trend in \protect\cite{bensby:13} is leveling off and continuing at the supersolar alpha values, with higher uncertainties. }
    \label{fig:alpha_fe_rm}
\end{figure*}

\subsection{MDF}

From Figure~\ref{fig:ns_all}, we find that the inner bulge MDF shows a bimodal distribution or the presence of two components. On the other hand, \citet{rich:2007,rich:12} did not find a bimodal metallicity distribution based on their detailed abundance analysis of 61 M giants located at ($l,b$)=($0\degr,-1\degr$), ($0\degr,-1\fdg75$), ($0\degr,-2\fdg75$) and Baade's window fields for the combined case as well as for individual fields. They find a mean metallicity of around $-0.2$\,dex with a typical dispersion of 0.15\,dex.  A different targeting strategy could be the reason for their absence of metal-rich stars.
\cite{babusiaux:14} also found no clear bimodality in the MDF of red clump stars located at the Northern field, ($l,b$) = ($0\degr,$+$1\degr$). However their spectra were of much lower spectral  resolution ($R \sim 6500$),  with low signal-to-noise ratio and the individual \feh\, measurement errors were larger, of the order of at least $0.2$\,dex.  In general, a larger sample would be required to assert bimodality; moreover, one would expect correlation with some additional physical parameter, such as radial velocity, proper motion, or composition.

The Baade Window field, one of the most observed Galactic bulge fields in terms of chemical abundances and global metallicities, reveals two components 
in their MDF (\citealt{schultheis:17,hill2011}) but these components are not established by other physical properties.
Bimodality in the MDF is also seen in recent studies using much larger samples of stars in the inner and outer bulge fields (\citealt{GIBSIII,Rojas17}). \cite{GIBSIII}, using red clump stars in GIBS survey, also at a lower spectral  resolution ($R \sim 6500$), using the Calcium II triplet (CaT) feature at $\sim9000\,$\AA, found a bimodality in their MDF for fields similar to $b$=$-1\degr$ and $b$=$-2\degr$, though we do not have enough number statistics in each field to reproduce and compare it with. These authors also found a bimodal MDFs for the outer bulge fields centered at ($l,b$) = ($0\degr,-6\degr$) and ($0\degr,-8\degr$) with the metal rich fraction getting lower as we move away from the Galactic mid-plane. Similar results  were obtained by \cite{Rojas17} using the derived metallicities of red clump stars from the Gaia-ESO survey for the  outer bulge fields.  Both the above mentioned studies assumed a Gaussian distribution for the bimodal components and used a Gaussian Mixture Model (GMM) to characterize them. Since our sample size is small and the GMM method requires a larger sample size, we prefer to use a simple KDE analysis for our MDF. 

  We show the entire MDF in Figure~\ref{fig:ns_all_vs_gc} (left panel) along with the distribution in red showing the MDF of the Galactic Center sample. We find that all GC stars do have  super-solar metallicities with a mean metallicity of $\sim$+0.3 dex and a tight dispersion of $\sim$ 0.10\,dex.  Our peak metallicity value at the GC is consistent with that from the GC field of APOGEE \citep{schultheis:15}. A large fraction of  super-solar
  stars has  been  found by \citet{Feldmeier-Krause2017}  based on low-resolution spectra in the nuclear star cluster. We will discuss in a forthcoming paper the chemical similarities as well as differences between
  the nuclear star cluster, the Galactic Center field population and the
  inner Galactic bulge.  Our high resolution study does not confirm the large number of stars with [Fe/H]$>+0.55$ dex found by \citet{Do:2015} and \citet{Feldmeier-Krause2017}.  We suspect that many abundances derived at low resolution using Bayesian code such as employed in this study will be too high at the metal rich end, perhaps due to a lack of appropriate templates.  For cool stars of high metallicity, there is substantial risk of blends, especially with molecules, that can result in spuriously high abundance estimates based on low resolution spectra.
  
 \citet{Feldmeier-Krause2017} do also report  stars at low metallicity $-1.6<$\feh$<0.8$\,dex, whereas we do not find any subsolar metallicty star in our GC sample. However, the APOGEE stars in \citet{schultheis:15} are mainly located at 
  $|b| > 0\fdg5$ and do not cover the actual GC region. We also want to stress
  that our GC sample consists of only nine stars. Clearly more observations are needed to confirm the absence of metal-poor  stars in the Galactic center.
 
We also find the median, over-all metallicity in each field to decrease as we move outward from the Galactic mid-plane. Note, however, that the form of the distribution changes too. This confirms the presence of a negative vertical over-all metallicity gradient in the inner bulge fields as it has been found for the low extinction fields of the outer bulge (\citealt{zoccali2008,gonzalez:11,ness:13}).       
There is one star in our $b$=+2\degr field that is very metal poor (\feh$=-1.8$) compared to our bulge sample. Our uncertainties, even at this low metallicity are definitely small.  Our estimated radial velocity of $-101$\,km\,s$^{-1}$ does not unequivocally suggest that it is a halo star passing through the bulge.  We note that \citet{schultheis:15} discovered, as mentioned above, a few stars with metallicities of this magnitude range , suggesting the presence of a metal-poor population.  Further, \citet{Do:2015} report several stars with [Fe/H]$<-1$ in the central cluster.  Confirmation of the nature of these stars will depend on proper motions and radial velocities.
  








\subsection{$\alpha$-element trends with [Fe/H]}

As mentioned in Section~\ref{result}, though we find the expected $\alpha$-abundance trend, the \sife\, abundances are lower than \mgfe\ abundances in the subsolar metallicity regime.
In Figure~\ref{fig:alpha_fe_rm}, we show the outer bulge ($|b|> 2^\circ$) $\alpha$-element trends seen for micro-lensed  dwarf sample in \cite{bensby:13} and for red giant branch stars in \cite{johnson:2014}. We plot each star with their individual uncertainties estimated for \sife\,,\mgfe\, and \feh, from the corresponding tables in \cite{bensby:13} and \cite{johnson:2014}. The $\alpha$ abundance trend with metallicity in \cite{rich:12} cover only a narrow range in metallicity of $\sim$-0.4 to 0\,dex. Hence, it is not possible to make a full metallicity range comparison with our trend though two of our sample fields coincide with theirs in the inner Galactic bulge.


We find that our \sife\ trend is consistent with the outer bulge trend of \citet{johnson:2014}
over the full metallicity ranges, apart from a systematic shift of the order of 0.1 dex. This shift is consistent within the 1-$\sigma$ uncertainty limits of both samples. Both trends seem to continuously decrease with metallicity, even in the supersolar regime. 
The \sife\ trend of \mbox{\cite{bensby:13}}  in the subsolar metallicity regime, lies consitently in between our and the \citet{johnson:2014} trend.  One significant difference is, however, that the supersolar \sife\ trend of \cite{bensby:13} saturates or levels off and continues at supersolar \sife\, values. Such a scenario is only possible if the rate of Type II supernovae explosions, that produce majority of $\alpha$ elements, increases sufficiently compared to the Type Ia supernovae explosions or if there is a metallicity dependent yield that increases the Si abundances dramatically,  both of which are unexpected. Another possibility is that our studies are considering different stellar populations. Also, generally it should be noted that at supersolar metallicities, there might be a problem with blending atomic or molecular lines, which are not known to exist or whose strengths are unknown. The lines used in abundance analyses are generally checked for blends, but if there are unknown blends which cannot be taken  into account, there is a tendency to overestimate the abundance of a spectra line, since the unknown blending line is ascribed to the element. Therefore, there is always a risk of overestimating the derived abundances in cool, metal rich stars.


Our \mgfe\, trend is consistent with the trends of both \cite{bensby:13} and \cite{johnson:2014} in the subsolar metallicity regime. Our trend over the entire metallicity ranges is especially consistent with that of \cite{johnson:2014}. Still the dispersion in our \mgfe\, measurements are higher than the comparison samples which may be due to the fact that there is only one spectral feature that can be used for the abundance determination, namely the group of lines at 21060\,\AA.  In the supersolar metallicity regime, our trend is consistent with that found by \cite{johnson:2014}, while the trend in \mbox{\cite{bensby:13}} is again leveling off and continuing at the supersolar \mgfe\, values.




\subsection{Kinematics}

\begin{figure}
    \includegraphics[width=\columnwidth]{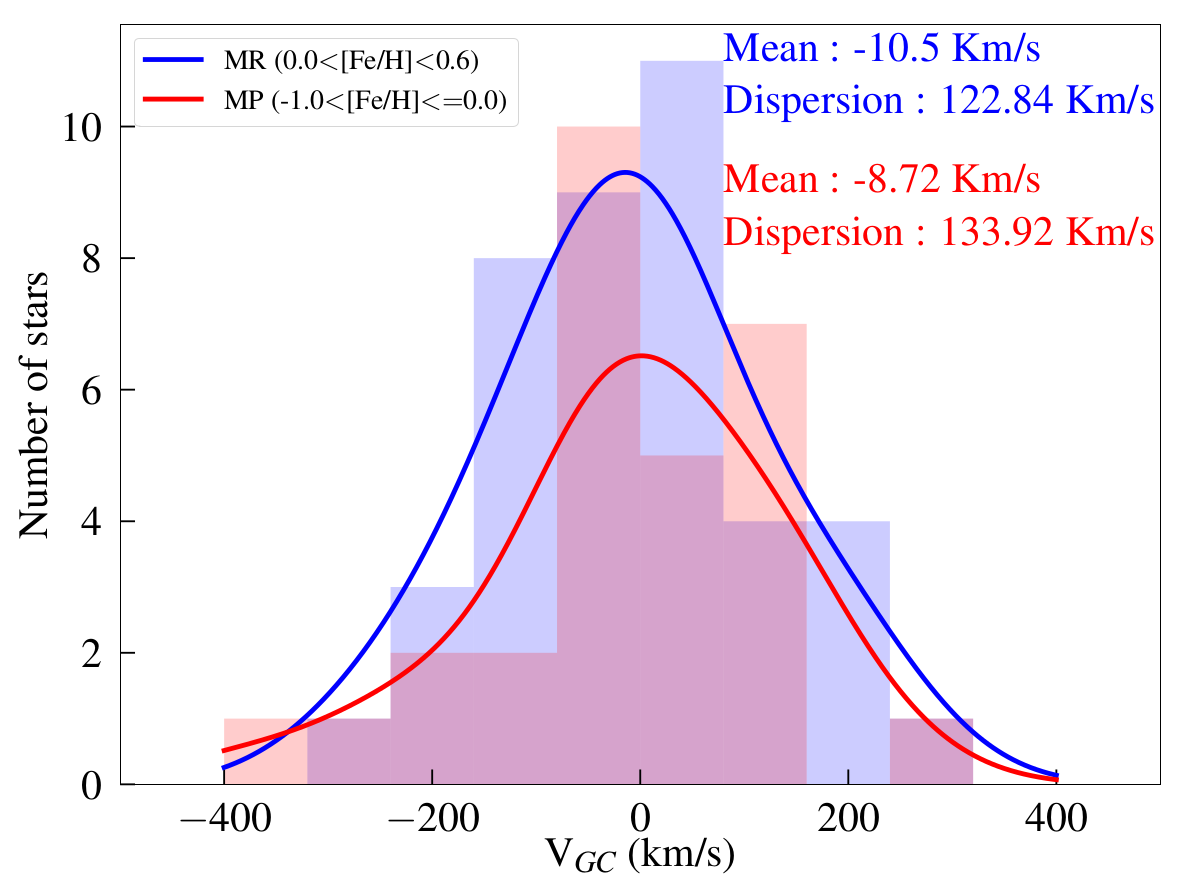}
    \caption{The Galactocentric radial velocity distribution for our entire sample divided into metal-poor (MP; red) and metal-rich (MR; blue). Overlaid KDEs of respective colors have similar bandwidth as their binsize. The mean velocity and dispersion for the two samples are listed in the figure.}
    \label{fig:vrad_mp_mr}
\end{figure}

\begin{figure*}
    \includegraphics[width=\columnwidth]{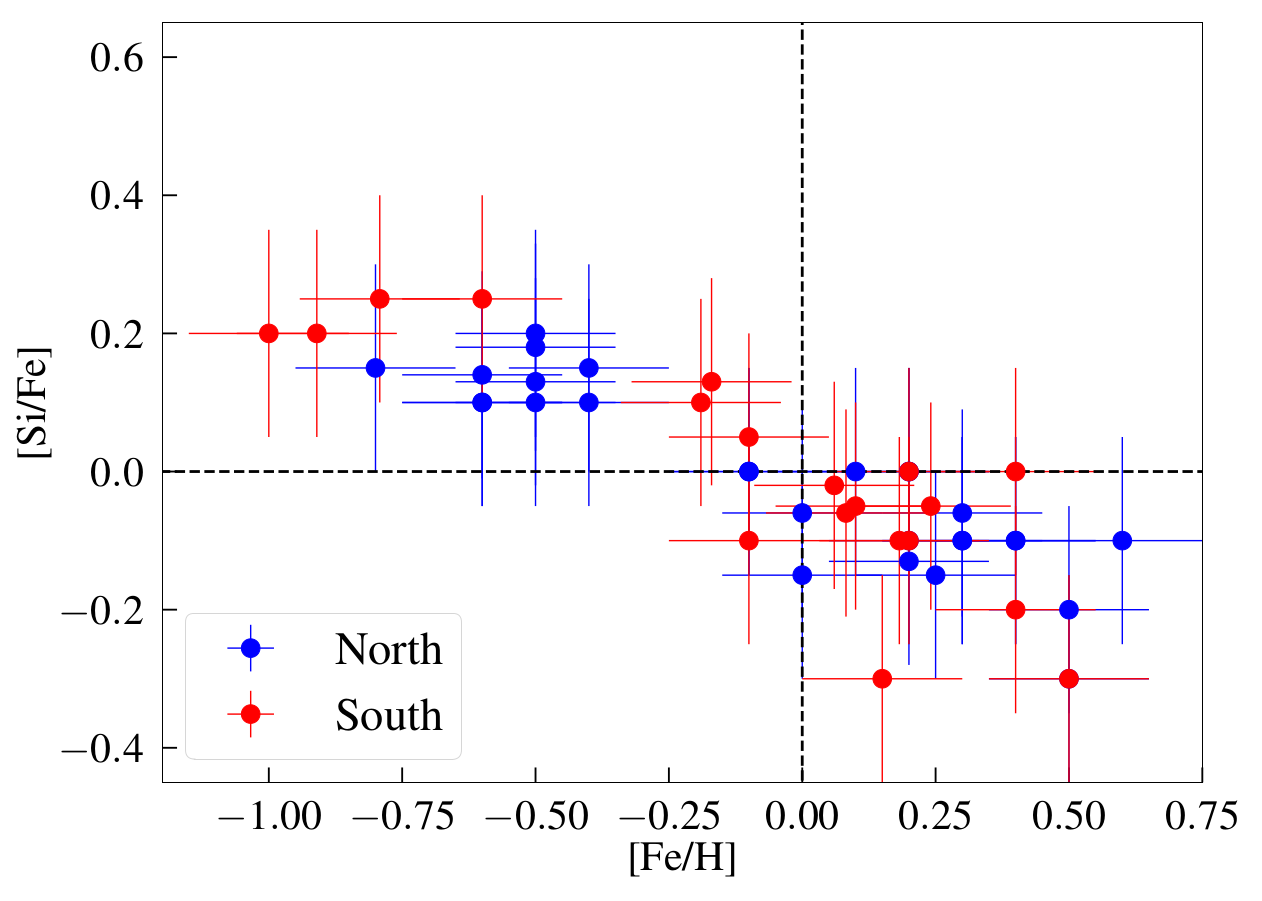}
    \includegraphics[width=\columnwidth]{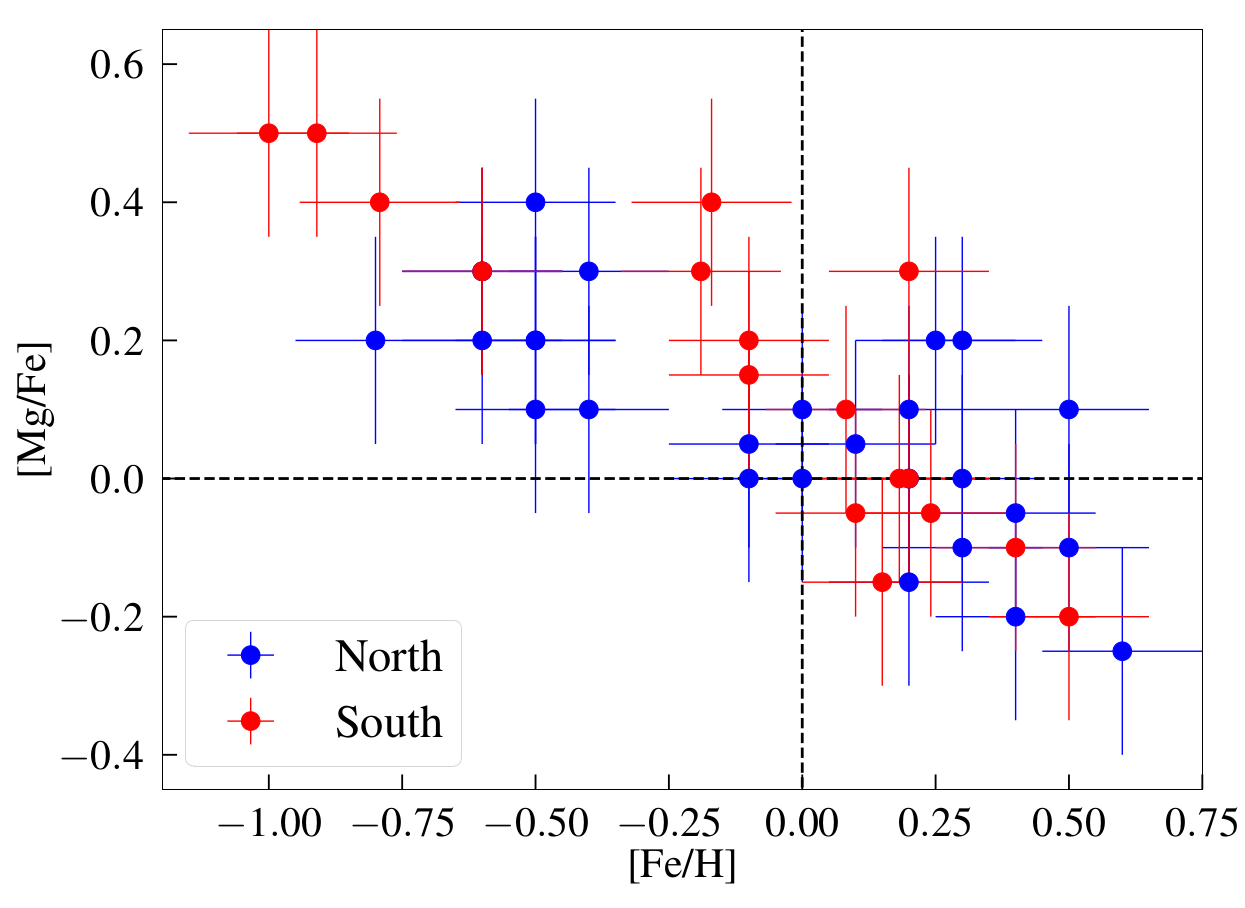}
    \caption{[Si/Fe] vs [Fe/H] (left) and [Mg/Fe] vs [Fe/H] (right) trends of stars at b=+1\degr, +2\degr (blue filled circles) and those at b=-1\degr, -2\degr (red filled circles). The error bars correspond to typical uncertainty of 0.15\,dex.}
    \label{fig:alpha_fe_ns}
\end{figure*}

The mean value of  the Galactocentric radial velocity that we have estimated for our entire sample within $|$b$|$<3\degr is consistent with the estimates from other studies using much larger sample of stars in the inner and outer bulge fields \citep{howard:08,kunder:12,ness:13a,zoccali:14,babusiaux:14}. The mean values of V$_{GC}$ estimated by BRAVA survey for fields at (l,b) $\sim$ (0\degr,-2\degr), (0\degr,-4\degr), (0\degr,-5\degr), (0\degr,-6\degr) and (0\degr,+4\degr) ranges from -8 to 14\,kms$^{-1}$ \citep{howard:08,kunder:12}. Our mean estimate of -12 kms$^{-1}$ is closer to the values of -9.7, -6.2 and -10.9\,kms$^{-1}$ found by \cite{ness:13a} for their fields at (l,b) = (0\degr,-5\degr), (0\degr,-7\fdg5) and (0\degr,-10\degr) respectively. \cite{babusiaux:14} also finds a negative mean V$_{GC}$ of -3\,kms$^{-1}$ for their sample at (l,b) = (0\degr,+1\degr). At the same time, the mean V$_{GC}$ surface in the longitude-latitude plane as derived from the GIBS Survey (Fig. 10 in \citealt{zoccali:14}\footnote{Assuming symmetry in b}) clearly showed that the mean V$_{GC}$ along the bulge minor axis is in the range of -10 to 0\,kms$^{-1}$. \citet{Valenti:2018} measured the mean Galactocentric radial velocity and dispersion from the low resolution MUSE spectra for bulge stars in the fields at (l,b) = (0\degr,+2\degr), (0\degr,-2\degr), (+1\degr,-1\degr) and (-1\degr,+2\degr). They find the mean Galactocentric radial velocity for stars in their fields corresponding to our bp2 and bm2 fields to be +9.8 and +19.2,kms$^{-1}$ respectively.

In the case of the  Galactocentric radial velocity dispersion, \cite{babusiaux:16}   compiled  radial velocity dispersion values along the bulge minor axis (l=0\degr, $|$b$|$<12\degr) from different literature studies. They showed that  the V$_{GC}$ dispersion follows a steep trend that goes down as we move away from the Galactic mid-plane. So, we compare our dispersion estimate with only those studies within $|$b$|$<2\degr. \citep{howard:08} estimated the velocity dispersion at (l,b) $\sim$ (0\degr,-2\degr) to be 126\,kms$^{-1}$, consistent with our estimate. A higher velocity dispersion of 145\,kms$^{-1}$ was calculated by \cite{babusiaux:14}. \cite{zoccali:14} also found a peak dispersion value closer to 140\,kms$^{-1}$ for the GIBS sample at (l,b) $\sim$ (0\degr,-1\degr) and (0\degr,-2\degr). \citet{Valenti:2018} confirmed this result in their study including fields at positive latitudes that were missing in GIBS, also finding a symmetric velocity dispersion about the Galactic plane. Any slight variation of our estimate from the rest of the studies can be attributed to our small sample size due to which we had to combine our entire sample.

\cite{babusiaux:16} has also shown (see their Fig. 4.) that the dispersion trends for the  metal-poor star  (-1<\feh\,<-0.5) and the metal-rich star sample (0<\feh\,<0.) are comparable within $|$b$|$<2\degr. For higher latitudes ($\rm |b| \geq 4\degr$) there is a strong decrease in velocity dispersion for metal-rich stars while it is flat for metal-poor stars.   By dividing the GIBS sample into metal-poor (-0.8<\feh\,<-0.1) and metal-rich (\feh\,>0.1) populations, \cite{GIBSIII} also found a similar trend. We also separated our limited sample into metal-poor (-1.0$\leq$\feh\,$\leq$0.0) and metal-rich (0.0<\feh\,$\leq$0.6) populations and estimated their dispersions as shown in Figure~\ref{fig:vrad_mp_mr}. We find similar velocity dispersions for the two populations which are close to the estimates from the above mentioned studies.  However,
we  suffer from a limited sample size and clearly more observations are needed.

\subsection{North-South Symmetry}

The North-South symmetry in bulge kinematics, namely mean Galactocentric radial velocity and dispersion, have already been shown for a range of longitudes by sampling outer bulge fields \citep{ness:13a,zoccali:14}. The rotation and dispersion symmetry along the major axis has been shown by \cite{ness:13a} using three fields at latitudes of -10\degr and +10\degr, and longitudes of -10\degr, -5\degr and +10\degr. Similar results has been obtained by \cite{zoccali:14}  using the GIBS survey fields at b = +4\fdg5 and -4\fdg7 covering a range of longitudes, $|$l$|$<8\degr. 

\cite{valenti:16} have shown, using the VVV PSF-fitting photometry, the increase of the  RC density distribution in the density profile of the inner bulge towards the Galactic plane. They have also shown symmetries in their  RC star counts in the Galactic mid-plane, at l = 0\degr, b = $\pm$1\degr, $\pm$2\degr, $\pm$3\degr and $\pm$4\degr. This
 suggests  a  North-South symmetry in kinematics and stellar density for the outer bulge and inner bulge regions respectively. We extend this investigation for the North-South symmetry to chemical abundances and the metallicity distribution function.

Four of our fields along the bulge minor axis are located at symmetrically opposite latitudes with respect to the Galactic mid-plane. In addition, we have carried out a consistent analysis of the spectral analysis  in both North and South fields. This is
crucial in order to explore the symmetry of the  MDF between Northern and Southern fields along the bulge minor axis. This is the first time such a study is being carried out, as previously most of the studies in the inner bulge have been concentrated to the southern latitudes. 

The Sun is located at a distance of 17$\pm$5\,pc above the Galactic mid-plane and at a distance of 8.2$\pm$0.1\,kpc from the Galactic center \citep{karim:17}. Based on this, we estimated very similar distance of $\sim$145 and 298\,pc, above the Galactic mid-plane along the bulge minor axis, corresponding to the North-South latitudes, $\pm$1\degr and $\pm$2\degr respectively. Thus we can safely assume that our  stars at the North-South latitudes in our sample are located symmetrically with respect to the Galactic mid-plane. 

We group the stars in the b=+1\degr, b=+2\degr fields together as the North field and those in the b=-1\degr, b=-2\degr fields as the South field, and plot their metallicity distributions in Figure~\ref{fig:ns_all_vs_gc} (right panel). 
The metal rich part of both the MDFs are very similar. The overlaid KDE estimate for  the North and South fields in the metal-rich regime are similar with approximate matching peaks. Keeping in mind the low number of stars and the abundance uncertainties, the KDE peak in the metal-poor regime for South field do, however, not seem to match with that for the Northern field due to the absence of stars with metallicities in the range of $-0.3$ to $-0.6$\,dex. The metallicity distribution for the Northern field  in the metal-poor regime show a peak in the KDE of  around $-0.5$\,dex. We want to stress 
that we are dealing here with low-number statistics; there are only four stars in the Southern fields with metallicities falling into the metal-poor group. Clearly a larger sample is needed. We can not reject the hypothesis of symmetry  based on the distribution of these few stars. It is too early to claim any difference.
Testing the full Northern and Southern MDF populations (only excluding the \feh=$-1.8$ star) with a Kolmogrov-smirnoff test, we arrive at a p-value of 0.90 suggesting that both distributions, nevertheless, come from the same parent population. Thus, we cannot claim any non-symmetry in the MDFs between Northern and Southern fields based on our data. Most likely the inner bulge is symmetric along the minor axis in its metallicity distributions.

We then proceed to carry out a similar exercise for the \sife\, vs \feh\, and \mgfe\, vs \feh\, trends, as shown in Figure~\ref{fig:alpha_fe_ns}. As mentioned in the above paragraph,  the absence of Southern field stars in the range of $-0.3$ to $-0.6$\,dex results in insignificant overlap between North-South samples in subsolar metallicity regime. Still, they a follow very similar continuous trend. On the other hand, there is significant overlap and consistency between the \sife\, and \mgfe\, values in the supersolar metallicity regime. Thus we cannot reject North-South symmetry in the $\alpha$-abundance trends based on our data, but we need more data to confirm it.


\section{Conclusions}

We have carried out a consistent spectroscopic analysis of 72 M giant stars in the inner Galactic bulge ($|$b$|$<3\degr) along the minor axis from the high resolution CRIRES spectra. Out of 72 stars, 9 stars are located in the Galactic center, 44 stars are at the Northern latitudes, while 28 at the Southern latitudes.  This is the first work to carry out such a systematic analysis in the inner bulge region, especially at symmetric fields along the bulge minor axis. 

We find a bimodal MDF in the inner bulge, following that reported for outer bulge samples, but a larger sample will be required for confirmation. Our GC sample is completely supersolar with a mean metallicity of $\rm 0.3\,dex$ but with no subsolar metallicities which could be due to our limited sample size.    We find that the median overall metallicities in each field to decrease as we move away from the Galactic mid-plane. This is in agreement with the negative vertical metallicity gradient found in outer bulge fields.   We find no stars with [Fe/H]$>+0.6$ dex.

The $\alpha$-element trends with metallicity show no gradient over the entire minor axis, from far out through the GC and going North. Our \sife\, and \mgfe\, trends with \feh\, are consistent with outer bulge trends for red giants, showing a gradual decreasing trend with increasing metallicity, as expected from chemical evolution models. We confirm that for the supersolar regime, the \sife\ and \mgfe\ abundances decrease as expected from chemical evolution models.

The mean Galactocentric radial velocity and dispersion for our entire sample is consistent within the  errors with the study of \citet{howard:08} and that of the GIBS survey. No difference in the velocity dispersion has been found between the metal-rich and the metal-poor sample.

Based on our investigation using b = $\pm$1\degr, $\pm$2\degr fields, we confirm the expected North-South symmetry in  the MDF as well as in the $\alpha$ abundances for the inner bulge region. These results imply that there is  chemical homogenity between the northern and the southern fields.

Although we report the largest and highest resolution abundance analysis of giants within 2 degrees of the Galactic center, the sample size remains relativley modest.
Upcoming instruments and surveys such as  CRIRES+ or MOONS will help to obtain a larger sample of stars in the inner bulge and towards the Galactic Center region. This will allow us to trace  the vertical metallicity gradient as well as the full MDF in the GC together with their chemical footprints.  It will also be vital to add superior kinematic data.   All of this can help in deciphering the formation/evolution scenario of the inner bulge as well as the relation of the inner bulge to the  Galactic Center and its nuclear star cluster.


\section*{Acknowledgements}
We wish to thank the anonymous referee for the comments that improved the quality of this paper. G.N and M.S. acknowledges the Programme National de Cosmologie et Galaxies (PNCG) of CNRS/INSU, France, for financial support. N.R. acknowledges support from the Swedish Research Council, VR (project number 621-2014-5640). N.R. acknowledges Funds from Kungl. Fysiografiska S\"allskapet i Lund. (Stiftelsen Walter Gyllenbergs fond and M\"arta och Erik Holmbergs donation). P.S.B. gratefully acknowledges the support of the Swedish Research Council. N.R. and P.S.B. acknowledge Funds from the project grant `The New Milky Way' from the Knut and Alice Wallenberg foundation. H.J. acknowledges support from the Crafoord Foundation, and Stiftelsen Olle Engkvist Byggm\"astare.




\bibliographystyle{mnras}
\bibliography{references} 

\begin{thebibliography}{}
\makeatletter
\relax
\def\mn@urlcharsother{\let\do\@makeother \do\$\do\&\do\#\do\^\do\_\do\%\do\~}
\def\mn@doi{\begingroup\mn@urlcharsother \@ifnextchar [ {\mn@doi@}
  {\mn@doi@[]}}
\def\mn@doi@[#1]#2{\def\@tempa{#1}\ifx\@tempa\@empty \href
  {http://dx.doi.org/#2} {doi:#2}\else \href {http://dx.doi.org/#2} {#1}\fi
  \endgroup}
\def\mn@eprint#1#2{\mn@eprint@#1:#2::\@nil}
\def\mn@eprint@arXiv#1{\href {http://arxiv.org/abs/#1} {{\tt arXiv:#1}}}
\def\mn@eprint@dblp#1{\href {http://dblp.uni-trier.de/rec/bibtex/#1.xml}
  {dblp:#1}}
\def\mn@eprint@#1:#2:#3:#4\@nil{\def\@tempa {#1}\def\@tempb {#2}\def\@tempc
  {#3}\ifx \@tempc \@empty \let \@tempc \@tempb \let \@tempb \@tempa \fi \ifx
  \@tempb \@empty \def\@tempb {arXiv}\fi \@ifundefined
  {mn@eprint@\@tempb}{\@tempb:\@tempc}{\expandafter \expandafter \csname
  mn@eprint@\@tempb\endcsname \expandafter{\@tempc}}}

\bibitem[\protect\citeauthoryear{{Abadi}, {Navarro}, {Steinmetz}  \&
  {Eke}}{{Abadi} et~al.}{2003}]{abadi:03}
{Abadi} M.~G.,  {Navarro} J.~F.,  {Steinmetz} M.,   {Eke} V.~R.,  2003, \mn@doi
  [\apj] {10.1086/375512}, \href
  {http://adsabs.harvard.edu/abs/2003ApJ...591..499A} {591, 499}

\bibitem[\protect\citeauthoryear{{Anstee} \& {O'Mara}}{{Anstee} \&
  {O'Mara}}{1991}]{abo1}
{Anstee} S.~D.,  {O'Mara} B.~J.,  1991, \mn@doi [\mnras]
  {10.1093/mnras/253.3.549}, \href
  {http://adsabs.harvard.edu/abs/1991MNRAS.253..549A} {253, 549}

\bibitem[\protect\citeauthoryear{{Athanassoula}, {Rodionov}, {Peschken}  \&
  {Lambert}}{{Athanassoula} et~al.}{2016}]{athanassoula_merge:16}
{Athanassoula} E.,  {Rodionov} S.~A.,  {Peschken} N.,   {Lambert} J.~C.,  2016,
  \mn@doi [\apj] {10.3847/0004-637X/821/2/90}, \href
  {http://adsabs.harvard.edu/abs/2016ApJ...821...90A} {821, 90}

\bibitem[\protect\citeauthoryear{{Babusiaux}}{{Babusiaux}}{2016}]{babusiaux:16}
{Babusiaux} C.,  2016, \mn@doi [\pasa] {10.1017/pasa.2016.1}, \href
  {http://adsabs.harvard.edu/abs/2016PASA...33...26B} {33, e026}

\bibitem[\protect\citeauthoryear{{Babusiaux} et~al.,}{{Babusiaux}
  et~al.}{2014}]{babusiaux:14}
{Babusiaux} C.,  et~al., 2014, \mn@doi [\aap] {10.1051/0004-6361/201323044},
  \href {http://adsabs.harvard.edu/abs/2014A%26A...563A..15B} {563, A15}

\bibitem[\protect\citeauthoryear{{Barbuy}, {Chiappini}  \& {Gerhard}}{{Barbuy}
  et~al.}{2018}]{Barbuy:2018}
{Barbuy} B.,  {Chiappini} C.,   {Gerhard} O.,  2018, preprint, \href
  {http://adsabs.harvard.edu/abs/2018arXiv180501142B} {} (\mn@eprint {arXiv}
  {1805.01142})

\bibitem[\protect\citeauthoryear{{Barklem} \& {O'Mara}}{{Barklem} \&
  {O'Mara}}{1998}]{abo2}
{Barklem} P.~S.,  {O'Mara} B.~J.,  1998, \mn@doi [\mnras]
  {10.1046/j.1365-8711.1998.01942.x}, \href
  {http://adsabs.harvard.edu/abs/1998MNRAS.300..863B} {300, 863}

\bibitem[\protect\citeauthoryear{{Barklem}, {Piskunov}  \& {O'Mara}}{{Barklem}
  et~al.}{2000}]{barklem:00}
{Barklem} P.~S.,  {Piskunov} N.,   {O'Mara} B.~J.,  2000, A\&AS, 142, 467

\bibitem[\protect\citeauthoryear{{Bensby}, {Yee}, {Feltzing}  et~al.}{{Bensby}
  et~al.}{2013}]{bensby:13}
{Bensby} T.,  {Yee} J.~C.,  {Feltzing} S.,   et~al., 2013, \mn@doi [\aap]
  {10.1051/0004-6361/201220678}, \href
  {http://adsabs.harvard.edu/abs/2013A%26A...549A.147B} {549, A147}

\bibitem[\protect\citeauthoryear{{Bressan}, {Marigo}, {Girardi}, {Salasnich},
  {Dal Cero}, {Rubele}  \& {Nanni}}{{Bressan} et~al.}{2012}]{bressan:12}
{Bressan} A.,  {Marigo} P.,  {Girardi} L.,  {Salasnich} B.,  {Dal Cero} C.,
  {Rubele} S.,   {Nanni} A.,  2012, \mn@doi [\mnras]
  {10.1111/j.1365-2966.2012.21948.x}, \href
  {http://adsabs.harvard.edu/abs/2012MNRAS.427..127B} {427, 127}

\bibitem[\protect\citeauthoryear{{Carr}, {Sellgren}  \& {Balachandran}}{{Carr}
  et~al.}{2000}]{carr2000}
{Carr} J.~S.,  {Sellgren} K.,   {Balachandran} S.~C.,  2000, \mn@doi [\apj]
  {10.1086/308340}, \href {http://adsabs.harvard.edu/abs/2000ApJ...530..307C}
  {530, 307}

\bibitem[\protect\citeauthoryear{{Civi{\v s}}, {Ferus}, {Chernov}  \&
  {Zanozina}}{{Civi{\v s}} et~al.}{2013}]{civis:13}
{Civi{\v s}} S.,  {Ferus} M.,  {Chernov} V.~E.,   {Zanozina} E.~M.,  2013,
  \mn@doi [\aap] {10.1051/0004-6361/201321052}, \href
  {http://adsabs.harvard.edu/abs/2013A%26A...554A..24C} {554, A24}

\bibitem[\protect\citeauthoryear{{Cunha}, {Sellgren}, {Smith}, {Ramirez},
  {Blum}  \& {Terndrup}}{{Cunha} et~al.}{2007}]{cunha2007}
{Cunha} K.,  {Sellgren} K.,  {Smith} V.~V.,  {Ramirez} S.~V.,  {Blum} R.~D.,
  {Terndrup} D.~M.,  2007, \mn@doi [\apj] {10.1086/521813}, \href
  {http://cdsads.u-strasbg.fr/abs/2007ApJ...669.1011C} {669, 1011}

\bibitem[\protect\citeauthoryear{{Debattista}, {Ness}, {Gonzalez}, {Freeman},
  {Zoccali}  \& {Minniti}}{{Debattista} et~al.}{2017}]{debattista:17}
{Debattista} V.~P.,  {Ness} M.,  {Gonzalez} O.~A.,  {Freeman} K.,  {Zoccali}
  M.,   {Minniti} D.,  2017, \mn@doi [\mnras] {10.1093/mnras/stx947}, \href
  {http://adsabs.harvard.edu/abs/2017MNRAS.469.1587D} {469, 1587}

\bibitem[\protect\citeauthoryear{{Demarque}, {Woo}, {Kim}  \& {Yi}}{{Demarque}
  et~al.}{2004}]{demarque:04}
{Demarque} P.,  {Woo} J.-H.,  {Kim} Y.-C.,   {Yi} S.~K.,  2004, \mn@doi [\apjs]
  {10.1086/424966}, \href {http://adsabs.harvard.edu/abs/2004ApJS..155..667D}
  {155, 667}

\bibitem[\protect\citeauthoryear{{Di Matteo}}{{Di Matteo}}{2016}]{dimatteo:16}
{Di Matteo} P.,  2016, \mn@doi [\pasa] {10.1017/pasa.2016.11}, \href
  {http://adsabs.harvard.edu/abs/2016PASA...33...27D} {33, e027}

\bibitem[\protect\citeauthoryear{Di~Matteo et~al.,}{Di~Matteo
  et~al.}{2014}]{dimatteo:14}
Di~Matteo P.,  et~al., 2014, Astronomy and Astrophysics, 567, A122

\bibitem[\protect\citeauthoryear{Di~Matteo et~al.,}{Di~Matteo
  et~al.}{2015}]{dimatteo:15}
Di~Matteo P.,  et~al., 2015, Astronomy and Astrophysics, 577, A1

\bibitem[\protect\citeauthoryear{{Do}, {Kerzendorf}, {Winsor}, {St{\o}stad},
  {Morris}, {Lu}  \& {Ghez}}{{Do} et~al.}{2015}]{Do:2015}
{Do} T.,  {Kerzendorf} W.,  {Winsor} N.,  {St{\o}stad} M.,  {Morris} M.~R.,
  {Lu} J.~R.,   {Ghez} A.~M.,  2015, \mn@doi [\apj]
  {10.1088/0004-637X/809/2/143}, \href
  {http://adsabs.harvard.edu/abs/2015ApJ...809..143D} {809, 143}

\bibitem[\protect\citeauthoryear{{Eggen}, {Lynden-Bell}  \& {Sandage}}{{Eggen}
  et~al.}{1962}]{eggen:62}
{Eggen} O.~J.,  {Lynden-Bell} D.,   {Sandage} A.~R.,  1962, \mn@doi [\apj]
  {10.1086/147433}, \href {http://adsabs.harvard.edu/abs/1962ApJ...136..748E}
  {136, 748}

\bibitem[\protect\citeauthoryear{{Feldmeier-Krause}, {Kerzendorf}, {Neumayer},
  {Sch{\"o}del}, {Nogueras-Lara}, {Do}, {de Zeeuw}  \&
  {Kuntschner}}{{Feldmeier-Krause} et~al.}{2017}]{Feldmeier-Krause2017}
{Feldmeier-Krause} A.,  {Kerzendorf} W.,  {Neumayer} N.,  {Sch{\"o}del} R.,
  {Nogueras-Lara} F.,  {Do} T.,  {de Zeeuw} P.~T.,   {Kuntschner} H.,  2017,
  \mn@doi [\mnras] {10.1093/mnras/stw2339}, \href
  {http://adsabs.harvard.edu/abs/2017MNRAS.464..194F} {464, 194}

\bibitem[\protect\citeauthoryear{{Fragkoudi}, {Di Matteo}, {Haywood},
  {G{\'o}mez}, {Combes}, {Katz}  \& {Semelin}}{{Fragkoudi}
  et~al.}{2017}]{fragkoudi:17}
{Fragkoudi} F.,  {Di Matteo} P.,  {Haywood} M.,  {G{\'o}mez} A.,  {Combes} F.,
  {Katz} D.,   {Semelin} B.,  2017, \mn@doi [\aap]
  {10.1051/0004-6361/201630244}, \href
  {http://adsabs.harvard.edu/abs/2017A%26A...606A..47F} {606, A47}

\bibitem[\protect\citeauthoryear{{Fragkoudi}, {Di Matteo}, {Haywood},
  {Schultheis}, {Khoperskov}, {G{\'o}mez}  \& {Combes}}{{Fragkoudi}
  et~al.}{2018}]{fragkoudi:18}
{Fragkoudi} F.,  {Di Matteo} P.,  {Haywood} M.,  {Schultheis} M.,  {Khoperskov}
  S.,  {G{\'o}mez} A.,   {Combes} F.,  2018, preprint, \href
  {http://adsabs.harvard.edu/abs/2018arXiv180200453F} {} (\mn@eprint {arXiv}
  {1802.00453})

\bibitem[\protect\citeauthoryear{{Frogel}, {Tiede}  \& {Kuchinski}}{{Frogel}
  et~al.}{1999}]{frogel:99}
{Frogel} J.~A.,  {Tiede} G.~P.,   {Kuchinski} L.~E.,  1999, \mn@doi [\aj]
  {10.1086/300832}, \href {http://adsabs.harvard.edu/abs/1999AJ....117.2296F}
  {117, 2296}

\bibitem[\protect\citeauthoryear{{Garc{\'{\i}}a P{\'e}rez}
  et~al.,}{{Garc{\'{\i}}a P{\'e}rez} et~al.}{2018}]{garciaperez:18}
{Garc{\'{\i}}a P{\'e}rez} A.~E.,  et~al., 2018, \mn@doi [\apj]
  {10.3847/1538-4357/aa9d88}, \href
  {http://adsabs.harvard.edu/abs/2018ApJ...852...91G} {852, 91}

\bibitem[\protect\citeauthoryear{{Gardner}, {Debattista}, {Robin},
  {V{\'a}squez}  \& {Zoccali}}{{Gardner} et~al.}{2014}]{gardner:14}
{Gardner} E.,  {Debattista} V.~P.,  {Robin} A.~C.,  {V{\'a}squez} S.,
  {Zoccali} M.,  2014, \mn@doi [\mnras] {10.1093/mnras/stt2430}, \href
  {http://adsabs.harvard.edu/abs/2014MNRAS.438.3275G} {438, 3275}

\bibitem[\protect\citeauthoryear{{Gonzalez} et~al.,}{{Gonzalez}
  et~al.}{2011}]{gonzalez:11}
{Gonzalez} O.~A.,  et~al., 2011, \mn@doi [\aap] {10.1051/0004-6361/201116548},
  \href {http://adsabs.harvard.edu/abs/2011A%26A...530A..54G} {530, A54}

\bibitem[\protect\citeauthoryear{{Gonzalez}, {Rejkuba}, {Zoccali}, {Valenti},
  {Minniti}, {Schultheis}, {Tobar}  \& {Chen}}{{Gonzalez}
  et~al.}{2012}]{gonzalez2012}
{Gonzalez} O.~A.,  {Rejkuba} M.,  {Zoccali} M.,  {Valenti} E.,  {Minniti} D.,
  {Schultheis} M.,  {Tobar} R.,   {Chen} B.,  2012, \mn@doi [\aap]
  {10.1051/0004-6361/201219222}, \href
  {http://adsabs.harvard.edu/abs/2012A%26A...543A..13G} {543, A13}

\bibitem[\protect\citeauthoryear{{Grevesse}, {Asplund}  \& {Sauval}}{{Grevesse}
  et~al.}{2007}]{solar:sme}
{Grevesse} N.,  {Asplund} M.,   {Sauval} A.~J.,  2007, \mn@doi [\ssr]
  {10.1007/s11214-007-9173-7}, \href
  {http://adsabs.harvard.edu/abs/2007SSRv..130..105G} {130, 105}

\bibitem[\protect\citeauthoryear{{Grieco}, {Matteucci}, {Ryde}, {Schultheis}
  \& {Uttenthaler}}{{Grieco} et~al.}{2015}]{grieco:15}
{Grieco} V.,  {Matteucci} F.,  {Ryde} N.,  {Schultheis} M.,   {Uttenthaler} S.,
   2015, \mn@doi [\mnras] {10.1093/mnras/stv729}, \href
  {http://adsabs.harvard.edu/abs/2015MNRAS.450.2094G} {450, 2094}

\bibitem[\protect\citeauthoryear{{Gustafsson}, {Edvardsson}, {Eriksson}
  et~al.}{{Gustafsson} et~al.}{2008}]{marcs:08}
{Gustafsson} B.,  {Edvardsson} B.,  {Eriksson} K.,   et~al., 2008, \aap, 486,
  951

\bibitem[\protect\citeauthoryear{{Hill} et~al.,}{{Hill}
  et~al.}{2011}]{hill2011}
{Hill} V.,  et~al., 2011, \mn@doi [\aap] {10.1051/0004-6361/200913757}, \href
  {http://adsabs.harvard.edu/abs/2011A%26A...534A..80H} {534, A80}

\bibitem[\protect\citeauthoryear{{Houdashelt}, {Bell}  \&
  {Sweigart}}{{Houdashelt} et~al.}{2000}]{houdashelt2000}
{Houdashelt} M.~L.,  {Bell} R.~A.,   {Sweigart} A.~V.,  2000, \mn@doi [\aj]
  {10.1086/301243}, \href {http://cdsads.u-strasbg.fr/abs/2000AJ....119.1448H}
  {119, 1448}

\bibitem[\protect\citeauthoryear{{Howard}, {Rich}, {Reitzel}, {Koch}, {De
  Propris}  \& {Zhao}}{{Howard} et~al.}{2008}]{howard:08}
{Howard} C.~D.,  {Rich} R.~M.,  {Reitzel} D.~B.,  {Koch} A.,  {De Propris} R.,
   {Zhao} H.,  2008, \mn@doi [\apj] {10.1086/592106}, \href
  {http://adsabs.harvard.edu/abs/2008ApJ...688.1060H} {688, 1060}

\bibitem[\protect\citeauthoryear{{Immeli}, {Samland}, {Gerhard}  \&
  {Westera}}{{Immeli} et~al.}{2004}]{immeli:04}
{Immeli} A.,  {Samland} M.,  {Gerhard} O.,   {Westera} P.,  2004, \mn@doi
  [\aap] {10.1051/0004-6361:20034282}, \href
  {http://adsabs.harvard.edu/abs/2004A%26A...413..547I} {413, 547}

\bibitem[\protect\citeauthoryear{Johnson, Rich, Kobayashi, Kunder  \&
  Koch}{Johnson et~al.}{2014}]{johnson:2014}
Johnson C.~I.,  Rich R.~M.,  Kobayashi C.,  Kunder A.,   Koch A.,  2014, THE
  ASTRONOMICAL JOURNAL, 148, 67

\bibitem[\protect\citeauthoryear{{Karim} \& {Mamajek}}{{Karim} \&
  {Mamajek}}{2017}]{karim:17}
{Karim} M.~T.,  {Mamajek} E.~E.,  2017, \mn@doi [\mnras]
  {10.1093/mnras/stw2772}, \href
  {http://adsabs.harvard.edu/abs/2017MNRAS.465..472K} {465, 472}

\bibitem[\protect\citeauthoryear{{K\"aufl}, {Ballester}, {Biereichel}
  et~al.}{{K\"aufl} et~al.}{2004}]{crires}
{K\"aufl} H.-U.,  {Ballester} P.,  {Biereichel} P.,   et~al., 2004, in
  {Moorwood} A.~F.~M.,  {Iye} M.,  eds,  Presented at the Society of
  Photo-Optical Instrumentation Engineers (SPIE) Conference Vol. 5492, Society
  of Photo-Optical Instrumentation Engineers (SPIE) Conference Series. pp
  1218--1227, \mn@doi{10.1117/12.551480}

\bibitem[\protect\citeauthoryear{{K{\"a}ufl}, {Amico}, {Ballester}
  et~al.}{{K{\"a}ufl} et~al.}{2006}]{crires2}
{K{\"a}ufl} H.~U.,  {Amico} P.,  {Ballester} P.,   et~al., 2006, The Messenger,
  126, 32

\bibitem[\protect\citeauthoryear{Kaulakys}{Kaulakys}{1985}]{kaulakys:85}
Kaulakys B.,  1985, Journal of Physics B: Atomic and Molecular Physics, 18,
  L167

\bibitem[\protect\citeauthoryear{Kaulakys}{Kaulakys}{1991}]{kaulakys:91}
Kaulakys B.,  1991, Journal of Physics B: Atomic, Molecular and Optical
  Physics, 24, L127

\bibitem[\protect\citeauthoryear{{Klein Gebbinck},   \& {Sforna}}{{Klein
  Gebbinck} et~al.}{2012}]{gasgano}
{Klein Gebbinck} M.,    {Sforna} D.,  2012, {VERY LARGE TELESCOPE: Gasgano
  User’s Manual}

\bibitem[\protect\citeauthoryear{{Kunder} et~al.,}{{Kunder}
  et~al.}{2012}]{kunder:12}
{Kunder} A.,  et~al., 2012, \mn@doi [\aj] {10.1088/0004-6256/143/3/57}, \href
  {http://adsabs.harvard.edu/abs/2012AJ....143...57K} {143, 57}

\bibitem[\protect\citeauthoryear{{Kupka}, {Piskunov}, {Ryabchikova}, {Stempels}
   \& {Weiss}}{{Kupka} et~al.}{1999}]{vald2}
{Kupka} F.,  {Piskunov} N.,  {Ryabchikova} T.~A.,  {Stempels} H.~C.,   {Weiss}
  W.~W.,  1999, \mn@doi [\aaps] {10.1051/aas:1999267}, \href
  {http://adsabs.harvard.edu/abs/1999A%26AS..138..119K} {138, 119}

\bibitem[\protect\citeauthoryear{{Kupka}, {Ryabchikova}, {Piskunov}, {Stempels}
   \& {Weiss}}{{Kupka} et~al.}{2000}]{vald4}
{Kupka} F.~G.,  {Ryabchikova} T.~A.,  {Piskunov} N.~E.,  {Stempels} H.~C.,
  {Weiss} W.~W.,  2000, Baltic Astronomy, \href
  {http://adsabs.harvard.edu/abs/2000BaltA...9..590K} {9, 590}

\bibitem[\protect\citeauthoryear{{Lawrence} et~al.,}{{Lawrence}
  et~al.}{2013}]{ukidss}
{Lawrence} A.,  et~al., 2013, VizieR Online Data Catalog, \href
  {http://adsabs.harvard.edu/abs/2013yCat.2319....0L} {2319, 0}

\bibitem[\protect\citeauthoryear{{Livingston} \& {Wallace}}{{Livingston} \&
  {Wallace}}{1991}]{solar_IR_atlas}
{Livingston} W.,  {Wallace} L.,  1991, {An atlas of the solar spectrum in the
  infrared from 1850 to 9000 cm-1 (1.1 to 5.4 micrometer)}.
NSO Technical Report, Tucson: National Solar Observatory, National Optical
  Astronomy Observatory, 1991

\bibitem[\protect\citeauthoryear{{Majewski} et~al.,}{{Majewski}
  et~al.}{2017}]{majewski:17}
{Majewski} S.~R.,  et~al., 2017, \mn@doi [\aj] {10.3847/1538-3881/aa784d},
  \href {http://adsabs.harvard.edu/abs/2017AJ....154...94M} {154, 94}

\bibitem[\protect\citeauthoryear{{Martinez-Valpuesta} \&
  {Gerhard}}{{Martinez-Valpuesta} \& {Gerhard}}{2011}]{valpuesta:11}
{Martinez-Valpuesta} I.,  {Gerhard} O.,  2011, \mn@doi [\apjl]
  {10.1088/2041-8205/734/1/L20}, \href
  {http://adsabs.harvard.edu/abs/2011ApJ...734L..20M} {734, L20}

\bibitem[\protect\citeauthoryear{{McWilliam} \& {Rich}}{{McWilliam} \&
  {Rich}}{1994}]{mcwilliam:94}
{McWilliam} A.,  {Rich} R.~M.,  1994, \mn@doi [\apjs] {10.1086/191954}, \href
  {http://adsabs.harvard.edu/abs/1994ApJS...91..749M} {91, 749}

\bibitem[\protect\citeauthoryear{{Moorwood}}{{Moorwood}}{2005}]{crires1}
{Moorwood} A.,  2005, in {K{\"a}ufl} H.~U.,  {Siebenmorgen} R.,   {Moorwood}
  A.~F.~M.,  eds, High Resolution Infrared Spectroscopy in Astronomy. p.~15

\bibitem[\protect\citeauthoryear{{Moorwood}, {Cuby}  \& {Lidman}}{{Moorwood}
  et~al.}{1998a}]{sofi}
{Moorwood} A.,  {Cuby} J.-G.,   {Lidman} C.,  1998a, The Messenger, \href
  {http://adsabs.harvard.edu/abs/1998Msngr..91....9M} {91, 9}

\bibitem[\protect\citeauthoryear{{Moorwood} et~al.,}{{Moorwood}
  et~al.}{1998b}]{isaac}
{Moorwood} A.,  et~al., 1998b, The Messenger, \href
  {http://esoads.eso.org/abs/1998Msngr..94....7M} {94, 7}

\bibitem[\protect\citeauthoryear{{Nataf}, {Udalski}, {Gould}, {Fouqu{\'e}}  \&
  {Stanek}}{{Nataf} et~al.}{2010}]{nataf:10}
{Nataf} D.~M.,  {Udalski} A.,  {Gould} A.,  {Fouqu{\'e}} P.,   {Stanek} K.~Z.,
  2010, \mn@doi [\apjl] {10.1088/2041-8205/721/1/L28}, \href
  {http://adsabs.harvard.edu/abs/2010ApJ...721L..28N} {721, L28}

\bibitem[\protect\citeauthoryear{{Ness} et~al.,}{{Ness}
  et~al.}{2013a}]{ness:13}
{Ness} M.,  et~al., 2013a, \mn@doi [\mnras] {10.1093/mnras/sts629}, \href
  {http://adsabs.harvard.edu/abs/2013MNRAS.430..836N} {430, 836}

\bibitem[\protect\citeauthoryear{{Ness} et~al.,}{{Ness}
  et~al.}{2013b}]{ness:13a}
{Ness} M.,  et~al., 2013b, \mn@doi [\mnras] {10.1093/mnras/stt533}, \href
  {http://adsabs.harvard.edu/abs/2013MNRAS.432.2092N} {432, 2092}

\bibitem[\protect\citeauthoryear{{Nishiyama}, {Tamura}, {Hatano}, {Kato},
  {Tanab{\'e}}, {Sugitani}  \& {Nagata}}{{Nishiyama}
  et~al.}{2009}]{nishiyama2009}
{Nishiyama} S.,  {Tamura} M.,  {Hatano} H.,  {Kato} D.,  {Tanab{\'e}} T.,
  {Sugitani} K.,   {Nagata} T.,  2009, \mn@doi [\apj]
  {10.1088/0004-637X/696/2/1407}, \href
  {http://adsabs.harvard.edu/abs/2009ApJ...696.1407N} {696, 1407}

\bibitem[\protect\citeauthoryear{{Omont} et~al.,}{{Omont}
  et~al.}{2003}]{omont2003}
{Omont} A.,  et~al., 2003, \mn@doi [\aap] {10.1051/0004-6361:20030437}, \href
  {http://adsabs.harvard.edu/abs/2003A%26A...403..975O} {403, 975}

\bibitem[\protect\citeauthoryear{{Osorio}, {Barklem}, {Lind}, {Belyaev},
  {Spielfiedel}, {Guitou}  \& {Feautrier}}{{Osorio} et~al.}{2015}]{osorio:15}
{Osorio} Y.,  {Barklem} P.~S.,  {Lind} K.,  {Belyaev} A.~K.,  {Spielfiedel} A.,
   {Guitou} M.,   {Feautrier} N.,  2015, \mn@doi [\aap]
  {10.1051/0004-6361/201525846}, \href
  {http://adsabs.harvard.edu/abs/2015A%26A...579A..53O} {579, A53}

\bibitem[\protect\citeauthoryear{{Pehlivan Rhodin}, {Hartman}, {Nilsson}  \&
  {J{\"o}nsson}}{{Pehlivan Rhodin} et~al.}{2017}]{pehlivan:mg}
{Pehlivan Rhodin} A.,  {Hartman} H.,  {Nilsson} H.,   {J{\"o}nsson} P.,  2017,
  \mn@doi [\aap] {10.1051/0004-6361/201629849}, \href
  {http://adsabs.harvard.edu/abs/2017A%26A...598A.102P} {598, A102}

\bibitem[\protect\citeauthoryear{{Persson}, {West}, {Carr}, {Sivaramakrishnan}
  \& {Murphy}}{{Persson} et~al.}{1992}]{persson:92}
{Persson} S.~E.,  {West} S.~C.,  {Carr} D.~M.,  {Sivaramakrishnan} A.,
  {Murphy} D.~C.,  1992, \mn@doi [\pasp] {10.1086/132979}, \href
  {http://adsabs.harvard.edu/abs/1992PASP..104..204P} {104, 204}

\bibitem[\protect\citeauthoryear{{Piskunov} \& {Valenti}}{{Piskunov} \&
  {Valenti}}{2017}]{SME:2017}
{Piskunov} N.,  {Valenti} J.~A.,  2017, \mn@doi [\aap]
  {10.1051/0004-6361/201629124}, \href
  {http://adsabs.harvard.edu/abs/2017A%26A...597A..16P} {597, A16}

\bibitem[\protect\citeauthoryear{{Piskunov}, {Kupka}, {Ryabchikova}, {Weiss}
  \& {Jeffery}}{{Piskunov} et~al.}{1995}]{vald}
{Piskunov} N.~E.,  {Kupka} F.,  {Ryabchikova} T.~A.,  {Weiss} W.~W.,
  {Jeffery} C.~S.,  1995, A\&AS, 112, 525

\bibitem[\protect\citeauthoryear{{Ramirez}, {Depoy}, {Frogel}, {Sellgren}  \&
  {Blum}}{{Ramirez} et~al.}{1997}]{ramirez1997}
{Ramirez} S.~V.,  {Depoy} D.~L.,  {Frogel} J.~A.,  {Sellgren} K.,   {Blum}
  R.~D.,  1997, \mn@doi [\aj] {10.1086/118356}, \href
  {http://adsabs.harvard.edu/abs/1997AJ....113.1411R} {113, 1411}

\bibitem[\protect\citeauthoryear{{Ram{\'{\i}}rez}, {Stephens}, {Frogel}  \&
  {DePoy}}{{Ram{\'{\i}}rez} et~al.}{2000a}]{ramirez:00}
{Ram{\'{\i}}rez} S.~V.,  {Stephens} A.~W.,  {Frogel} J.~A.,   {DePoy} D.~L.,
  2000a, \mn@doi [AJ] {10.1086/301466}, \href
  {http://adsabs.harvard.edu/abs/2000AJ....120..833R} {120, 833}

\bibitem[\protect\citeauthoryear{{Ram{\'{\i}}rez}, {Sellgren}, {Carr},
  {Balachandran}, {Blum}, {Terndrup}  \& {Steed}}{{Ram{\'{\i}}rez}
  et~al.}{2000b}]{ramirez2000}
{Ram{\'{\i}}rez} S.~V.,  {Sellgren} K.,  {Carr} J.~S.,  {Balachandran} S.~C.,
  {Blum} R.,  {Terndrup} D.~M.,   {Steed} A.,  2000b, \mn@doi [\apj]
  {10.1086/309022}, \href {http://adsabs.harvard.edu/abs/2000ApJ...537..205R}
  {537, 205}

\bibitem[\protect\citeauthoryear{{Rich}, {Origlia}  \& {Valenti}}{{Rich}
  et~al.}{2007a}]{rich:2007}
{Rich} R.~M.,  {Origlia} L.,   {Valenti} E.,  2007a, \mn@doi [\apj]
  {10.1086/521440}, \href {http://adsabs.harvard.edu/abs/2007ApJ...665L.119R}
  {665, L119}

\bibitem[\protect\citeauthoryear{{Rich}, {Origlia}  \& {Valenti}}{{Rich}
  et~al.}{2007b}]{rich:07}
{Rich} R.~M.,  {Origlia} L.,   {Valenti} E.,  2007b, \mn@doi [\apjl]
  {10.1086/521440}, \href {http://adsabs.harvard.edu/abs/2007ApJ...665L.119R}
  {665, L119}

\bibitem[\protect\citeauthoryear{{Rich}, {Origlia}  \& {Valenti}}{{Rich}
  et~al.}{2012}]{rich:12}
{Rich} R.~M.,  {Origlia} L.,   {Valenti} E.,  2012, \mn@doi [\apj]
  {10.1088/0004-637X/746/1/59}, \href
  {http://adsabs.harvard.edu/abs/2012ApJ...746...59R} {746, 59}

\bibitem[\protect\citeauthoryear{{Rich}, {Ryde}, {Thorsbro}, {Fritz},
  {Schultheis}, {Origlia}  \& {J{\"o}nsson}}{{Rich} et~al.}{2017}]{rich:17}
{Rich} R.~M.,  {Ryde} N.,  {Thorsbro} B.,  {Fritz} T.~K.,  {Schultheis} M.,
  {Origlia} L.,   {J{\"o}nsson} H.,  2017, \mn@doi [\aj]
  {10.3847/1538-3881/aa970a}, \href
  {http://adsabs.harvard.edu/abs/2017AJ....154..239R} {154, 239}

\bibitem[\protect\citeauthoryear{{Rojas-Arriagada} et~al.,}{{Rojas-Arriagada}
  et~al.}{2014}]{rojas:15}
{Rojas-Arriagada} A.,  et~al., 2014, \mn@doi [\aap]
  {10.1051/0004-6361/201424121}, \href
  {http://adsabs.harvard.edu/abs/2014A%26A...569A.103R} {569, A103}

\bibitem[\protect\citeauthoryear{{Rojas-Arriagada} et~al.,}{{Rojas-Arriagada}
  et~al.}{2017}]{Rojas17}
{Rojas-Arriagada} A.,  et~al., 2017, \mn@doi [\aap]
  {10.1051/0004-6361/201629160}, \href
  {http://adsabs.harvard.edu/abs/2017A%26A...601A.140R} {601, A140}

\bibitem[\protect\citeauthoryear{{Ryabchikova}, {Piskunov}, {Kupka}  \&
  {Weiss}}{{Ryabchikova} et~al.}{1997}]{vald3}
{Ryabchikova} T.~A.,  {Piskunov} N.~E.,  {Kupka} F.,   {Weiss} W.~W.,  1997,
  Baltic Astronomy, \href {http://cdsads.u-strasbg.fr/abs/1997BaltA...6..244R}
  {6, 244}

\bibitem[\protect\citeauthoryear{{Ryabchikova}, {Piskunov}, {Kurucz},
  {Stempels}, {Heiter}, {Pakhomov}  \& {Barklem}}{{Ryabchikova}
  et~al.}{2015}]{vald5}
{Ryabchikova} T.,  {Piskunov} N.,  {Kurucz} R.~L.,  {Stempels} H.~C.,  {Heiter}
  U.,  {Pakhomov} Y.,   {Barklem} P.~S.,  2015, \mn@doi [\physscr]
  {10.1088/0031-8949/90/5/054005}, \href
  {http://adsabs.harvard.edu/abs/2015PhyS...90e4005R} {90, 054005}

\bibitem[\protect\citeauthoryear{{Ryde} \& {Schultheis}}{{Ryde} \&
  {Schultheis}}{2015}]{ryde_schultheis:15}
{Ryde} N.,  {Schultheis} M.,  2015, \mn@doi [\aap]
  {10.1051/0004-6361/201424486}, \href
  {http://adsabs.harvard.edu/abs/2015A%26A...573A..14R} {573, A14}

\bibitem[\protect\citeauthoryear{{Ryde}, {Schultheis}, {Grieco}, {Matteucci},
  {Rich}  \& {Uttenthaler}}{{Ryde} et~al.}{2016a}]{ryde:16}
{Ryde} N.,  {Schultheis} M.,  {Grieco} V.,  {Matteucci} F.,  {Rich} R.~M.,
  {Uttenthaler} S.,  2016a, \mn@doi [AJ] {10.3847/0004-6256/151/1/1}, \href
  {http://adsabs.harvard.edu/abs/2016AJ....151....1R} {151, 1}

\bibitem[\protect\citeauthoryear{{Ryde}, {Fritz}, {Rich}, {Thorsbro},
  {Schultheis}, {Origlia}  \& {Chatzopoulos}}{{Ryde}
  et~al.}{2016b}]{ryde:16:nsc}
{Ryde} N.,  {Fritz} T.~K.,  {Rich} R.~M.,  {Thorsbro} B.,  {Schultheis} M.,
  {Origlia} L.,   {Chatzopoulos} S.,  2016b, \mn@doi [\apj]
  {10.3847/0004-637X/831/1/40}, \href
  {http://adsabs.harvard.edu/abs/2016ApJ...831...40R} {831, 40}

\bibitem[\protect\citeauthoryear{{Scannapieco} \& {Tissera}}{{Scannapieco} \&
  {Tissera}}{2003}]{scannapieco:03}
{Scannapieco} C.,  {Tissera} P.~B.,  2003, \mn@doi [\mnras]
  {10.1046/j.1365-8711.2003.06133.x}, \href
  {http://adsabs.harvard.edu/abs/2003MNRAS.338..880S} {338, 880}

\bibitem[\protect\citeauthoryear{{Schultheis} et~al.,}{{Schultheis}
  et~al.}{2014}]{schultheis:14}
{Schultheis} M.,  et~al., 2014, preprint, \href
  {http://adsabs.harvard.edu/abs/2014arXiv1405.0503S} {} (\mn@eprint {arXiv}
  {1405.0503})

\bibitem[\protect\citeauthoryear{{Schultheis} et~al.,}{{Schultheis}
  et~al.}{2015}]{schultheis:15}
{Schultheis} M.,  et~al., 2015, \mn@doi [\aap] {10.1051/0004-6361/201527027},
  \href {http://adsabs.harvard.edu/abs/2015A%26A...584A..45S} {584, A45}

\bibitem[\protect\citeauthoryear{{Schultheis}, {Ryde}  \&
  {Nandakumar}}{{Schultheis} et~al.}{2016}]{schultheis:16}
{Schultheis} M.,  {Ryde} N.,   {Nandakumar} G.,  2016, \mn@doi [\aap]
  {10.1051/0004-6361/201628266}, \href
  {http://adsabs.harvard.edu/abs/2016A%26A...590A...6S} {590, A6}

\bibitem[\protect\citeauthoryear{{Schultheis} et~al.,}{{Schultheis}
  et~al.}{2017}]{schultheis:17}
{Schultheis} M.,  et~al., 2017, \mn@doi [\aap] {10.1051/0004-6361/201630154},
  \href {http://adsabs.harvard.edu/abs/2017A%26A...600A..14S} {600, A14}

\bibitem[\protect\citeauthoryear{{Shen}, {Rich}, {Kormendy}, {Howard}, {De
  Propris}  \& {Kunder}}{{Shen} et~al.}{2010}]{shen:10}
{Shen} J.,  {Rich} R.~M.,  {Kormendy} J.,  {Howard} C.~D.,  {De Propris} R.,
  {Kunder} A.,  2010, \mn@doi [\apjl] {10.1088/2041-8205/720/1/L72}, \href
  {http://adsabs.harvard.edu/abs/2010ApJ...720L..72S} {720, L72}

\bibitem[\protect\citeauthoryear{{Smith} et~al.,}{{Smith}
  et~al.}{2013}]{smith:13}
{Smith} V.~V.,  et~al., 2013, \mn@doi [\apj] {10.1088/0004-637X/765/1/16},
  \href {http://adsabs.harvard.edu/abs/2013ApJ...765...16S} {765, 16}

\bibitem[\protect\citeauthoryear{{Smoker}}{{Smoker}}{2007}]{crires:manual}
{Smoker} J.,  2007, {Very Large Telescope Paranal Science Operations CRIRES
  User Manual}

\bibitem[\protect\citeauthoryear{{Smoker}, {Valenti}, {Asmus}, {Birstow},
  {Smette}, {Hilker}, {Wolff}  \& {Jung}}{{Smoker} et~al.}{2012}]{crires:cook}
{Smoker} J.,  {Valenti} E.,  {Asmus} D.,  {Birstow} P.,  {Smette} A.,  {Hilker}
  M.,  {Wolff} B.,   {Jung} Y.,  2012, {Very Large Telescope Paranal Science
  Operations CRIRES data reduction cookbook}

\bibitem[\protect\citeauthoryear{{Sneden}, {Lucatello}, {Ram}, {Brooke}  \&
  {Bernath}}{{Sneden} et~al.}{2014}]{sneden:14}
{Sneden} C.,  {Lucatello} S.,  {Ram} R.~S.,  {Brooke} J.~S.~A.,   {Bernath} P.,
   2014, \mn@doi [\apjs] {10.1088/0067-0049/214/2/26}, \href
  {http://adsabs.harvard.edu/abs/2014ApJS..214...26S} {214, 26}

\bibitem[\protect\citeauthoryear{{Tody}}{{Tody}}{1993}]{IRAF}
{Tody} D.,  1993, in {Hanisch} R.~J.,  {Brissenden} R.~J.~V.,   {Barnes} J.,
  eds, ASP Conf. Ser. 52: Astronomical Data Analysis Software and Systems II.
  p.~173

\bibitem[\protect\citeauthoryear{{Valenti} \& {Piskunov}}{{Valenti} \&
  {Piskunov}}{1996}]{sme}
{Valenti} J.~A.,  {Piskunov} N.,  1996, \aaps, \href
  {http://adsabs.harvard.edu/abs/1996A%26AS..118..595V} {118, 595}

\bibitem[\protect\citeauthoryear{{Valenti} \& {Piskunov}}{{Valenti} \&
  {Piskunov}}{2012}]{sme_code}
{Valenti} J.~A.,  {Piskunov} N.,  2012, {SME: Spectroscopy Made Easy}
  (\mn@eprint {ascl} {1202.013})

\bibitem[\protect\citeauthoryear{{Valenti} et~al.,}{{Valenti}
  et~al.}{2016}]{valenti:16}
{Valenti} E.,  et~al., 2016, \mn@doi [\aap] {10.1051/0004-6361/201527500},
  \href {http://adsabs.harvard.edu/abs/2016A%26A...587L...6V} {587, L6}

\bibitem[\protect\citeauthoryear{{Valenti} et~al.,}{{Valenti}
  et~al.}{2018}]{Valenti:2018}
{Valenti} E.,  et~al., 2018, preprint, \href
  {http://adsabs.harvard.edu/abs/2018arXiv180500275V} {} (\mn@eprint {arXiv}
  {1805.00275})

\bibitem[\protect\citeauthoryear{{Wegg} \& {Gerhard}}{{Wegg} \&
  {Gerhard}}{2013}]{wegg:13}
{Wegg} C.,  {Gerhard} O.,  2013, \mn@doi [\mnras] {10.1093/mnras/stt1376},
  \href {http://adsabs.harvard.edu/abs/2013MNRAS.435.1874W} {435, 1874}

\bibitem[\protect\citeauthoryear{{Weiland} et~al.,}{{Weiland}
  et~al.}{1994}]{weiland:94}
{Weiland} J.~L.,  et~al., 1994, \mn@doi [\apjl] {10.1086/187315}, \href
  {http://adsabs.harvard.edu/abs/1994ApJ...425L..81W} {425, L81}

\bibitem[\protect\citeauthoryear{{Zoccali}, {Hill}, {Lecureur}, {Barbuy},
  {Renzini}, {Minniti}, {G{\'o}mez}  \& {Ortolani}}{{Zoccali}
  et~al.}{2008}]{zoccali2008}
{Zoccali} M.,  {Hill} V.,  {Lecureur} A.,  {Barbuy} B.,  {Renzini} A.,
  {Minniti} D.,  {G{\'o}mez} A.,   {Ortolani} S.,  2008, \mn@doi [\aap]
  {10.1051/0004-6361:200809394}, \href
  {http://cdsads.u-strasbg.fr/abs/2008A%26A...486..177Z} {486, 177}

\bibitem[\protect\citeauthoryear{{Zoccali} et~al.,}{{Zoccali}
  et~al.}{2014}]{zoccali:14}
{Zoccali} M.,  et~al., 2014, \mn@doi [\aap] {10.1051/0004-6361/201323120},
  \href {http://adsabs.harvard.edu/abs/2014A%26A...562A..66Z} {562, A66}

\bibitem[\protect\citeauthoryear{{Zoccali} et~al.,}{{Zoccali}
  et~al.}{2017}]{GIBSIII}
{Zoccali} M.,  et~al., 2017, \mn@doi [\aap] {10.1051/0004-6361/201629805},
  \href {http://adsabs.harvard.edu/abs/2017A%26A...599A..12Z} {599, A12}

\bibitem[\protect\citeauthoryear{{van Loon} et~al.,}{{van Loon}
  et~al.}{2003}]{vanloon2003}
{van Loon} J.~T.,  et~al., 2003, \mn@doi [\mnras]
  {10.1046/j.1365-8711.2003.06134.x}, \href
  {http://adsabs.harvard.edu/abs/2003MNRAS.338..857V} {338, 857}

\makeatother
\end{thebibliography}



\bsp	
\label{lastpage}
\end{document}